\RequirePackage{etoolbox}
\csdef{input@path}{{style/}{graphics/}}
\documentclass[seceqn,aap,MSNbibl,dvips]{arximspdf}
\usepackage{mathbh}

\doi{10.1214/14-AAP1012} 
\volume{25}
\issue{2}
\pubyear{2015}
\firstpage{860}
\lastpage{897}

\makeatletter

\newcommand{\rrVert}{\Vert}
\newcommand{\rrvert}{\vert}
\newcommand{\llVert}{\Vert}
\newcommand{\llvert}{\vert}

\newtheorem{theorem}{Theorem}[section]
\newtheorem{lem}[theorem]{Lemma}
\newtheorem{prop}[theorem]{Proposition}

\newproclaim{rk}[theorem]{Remark}
\newproclaim{defi}[theorem]{Definition}

\newcommand{\E}{\mathbb{E}}
\newcommand{\rr}{{\mathbb{R}}}
\newcommand{\dd}{{\mathbb{D}}}

\newcommand{\cS}{\mathcal{S}}
\newcommand{\cL}{\mathcal{L}}
\newcommand{\cM}{\mathcal{M}}
\newcommand{\cP}{\mathcal{P}}
\newcommand{\cF}{\mathcal{F}}
\newcommand{\cB}{\mathcal{B}}

\newcommand{\sg}{\operatorname{sg}}
\newcommand{\supp}{\operatorname{Supp}}
\newcommand{\lip}{\operatorname{Lip}}
\renewcommand{\Re}{\operatorname{Re}}

\newcommand{\indiq}{\mathbh{1}}
\newcommand{\sm}{{s-}}

\renewcommand{\epsilon}{\varepsilon}

\newcommand{\fracc}[2]{{#1}/(#2)}
\newcommand{\MP}{\operatorname{MP}}
\makeatother

\begin{document}
\begin{frontmatter}

\title{Finiteness of entropy for the homogeneous Boltzmann equation with measure initial condition}
\runtitle{Homogeneous Boltzmann equations}

\begin{aug}
\author[A]{\fnms{Nicolas}~\snm{Fournier}\corref{}\ead[label=e1]{nicolas.fournier@upmc.fr}}
\runauthor{N. Fournier}
\affiliation{Universit\'e Pierre et Marie Curie}
\address[A]{Laboratoire de Probabilit\'es et Mod\`eles Al\'eatoires\\
UMR 7599\\
Universit\'e Pierre et Marie Curie\\
Case 188, 4 place Jussieu\\
F-75252 Paris Cedex 5\\
France\\
\printead{e1}} 
\end{aug}

\received{\smonth{3} \syear{2013}}

%
\begin{abstract}
We consider the $3D$
spatially homogeneous Boltzmann equation for (true) hard and moderately
soft potentials.
We assume that the initial condition is a probability measure with
finite energy and is not
a Dirac mass.
For hard potentials, we prove that any reasonable weak solution
immediately belongs to some Besov space.
For moderately soft potentials, we assume additionally that the initial
condition has a moment of sufficiently high order ($8$ is enough) and
prove the existence of a solution that immediately belongs to some
Besov space.
The considered solutions thus instantaneously become functions with a
finite entropy.
We also prove that in any case, any weak solution is immediately
supported by $\rr^3$.
\end{abstract}

%
\begin{keyword}[class=AMS]
\kwd{82C40}
\kwd{60J75}
\kwd{60H30}
\end{keyword}
\begin{keyword}
\kwd{Kinetic equations}
\kwd{regularization}
\kwd{absolute continuity}
\kwd{entropy}
\kwd{Besov spaces}
\end{keyword}

\end{frontmatter}

\section{Introduction and results}

\subsection{The Boltzmann equation}

We consider a spatially homogeneous gas modeled by the Boltzmann equation:
the density $f_t(v)$ of particles with velocity $v\in\rr^3$ at time
$t\geq0$ solves
%
\begin{equation}
\label{be} \hspace*{20pt}\partial_t f_t(v) = \int
_{\rr^3}dv_* \int_{{\mathbb{S}^2}}\,d\sigma B\bigl(
\llvert v-v_*\rrvert ,\cos \theta\bigr) \bigl[f_t\bigl(v'
\bigr)f_t\bigl(v'_*\bigr) -f_t(v)f_t(v_*)
\bigr],
\end{equation}
where
%
\begin{eqnarray}
\label{vprime} v'&=&\frac{v+v_*}{2} + \frac{\llvert v-v_*\rrvert }{2}\sigma,
\qquad v'_*=\frac{v+v_*}{2} -\frac{\llvert v-v_*\rrvert }{2}\sigma\quad
\mbox{and}
\nonumber
\\[-8pt]
\\[-8pt]
\cos\theta&=& \biggl\langle\frac{v-v_*}{\llvert v-v_*\rrvert }, \sigma \biggr\rangle.
\nonumber
\end{eqnarray}
The cross section $B(\llvert v-v_*\rrvert ,\cos\theta)\geq0$
depends on the type of interaction
between particles. We refer to the book of Cercignani \cite{c} for a
physical reference on the Boltzmann equation and to the
review papers of Villani \cite{v:h} and Alexandre \cite{a}
for many details on what is known from the mathematical point of view.
Conservation of mass, momentum and kinetic energy
hold for reasonable solutions,
and we classically may assume without loss of generality that
$\int_{\rr^3}f_0(v) \,dv=1$.

\subsection{Assumptions}

We will assume that for some $\gamma\in(-1,1)$, some $\nu\in(0,1)$
with $\gamma+\nu>0$,
some measurable function $b\dvtx  (0,\pi] \mapsto\rr_+$,
{\renewcommand{\theequation}{$A_{\gamma,\nu}$}
%
\begin{equation}
\label{aaa} \hspace*{25pt}\cases{\displaystyle B\bigl(\llvert v-v_*\rrvert ,
\cos\theta\bigr)\sin\theta= \llvert v-v_*\rrvert ^\gamma b(\theta),
\cr
\displaystyle \exists0<c_0<C_0, \qquad\forall\theta
\in(0,\pi/2],\qquad c_0 \theta^{-1-\nu}\leq b(\theta)\leq
C_0 \theta^{-1-\nu},
\cr
\displaystyle \forall\theta\in(\pi/2,
\pi],\qquad b(\theta)=0. } %
\end{equation}}
As noted in the introduction of \cite{advw},
this last assumption ($b=0$ on $(\pi/2,\pi]$)
is not a restriction since we can always reduce to this case by a
symmetry argument.
When particles collide by pairs due to a repulsive force
proportional to $1/r^s$ for some $s> 2$, then (\ref{aaa})
holds with
$\gamma=(s-5)/(s-1)$ and $\nu=2/(s-1)$.
Thus our study includes the case of hard potentials ($s>5$), Maxwell
molecules ($s=5$)
and moderately soft potentials [$s\in(3,5)$].

\setcounter{equation}{2}

\subsection{Functional spaces}\label{fs}

Let us introduce all the functional spaces we will use in this paper:
\begin{itemize}
\item $\cM(\rr^d)$ is the set of nonnegative finite measures on
$\rr^d$.

\item $\cP(\rr^d)$ is the set of probability measures on $\rr^d$.

\item $\cP_p(\rr^d)$ is the set of all $f\in\cP(\rr^d)$ such that
$m_p(f):=\int_{\rr^d}\llvert v\rrvert ^p f(dv)<\infty$.

\item $\lip_b(\rr^d)$ is the set of bounded globally
Lipschitz-continuous functions.

\item $C_b(\rr^d)$ is the set of bounded continuous functions.

\item $C_0(\rr^d)$ is the set of continuous functions vanishing
at infinity.

\item $C^1_c(\rr^d)$ is the set of compactly supported $C^1$ functions.

\item For $\alpha\in(0,1)$, $C^\alpha_b(\rr^d)$ is the set of all
functions $g$ such that
\[
\llVert g \rrVert _{C^\alpha_b(\rr^d)} := \sup_{x\in\rr^d}\bigl\llvert
g(x)\bigr\rrvert + \sup_{x,y\in\rr
^d, x\ne y}\frac{\llvert g(x)-g(y)\rrvert }{\llvert x-y\rrvert ^\alpha}<\infty.
\]

\item $L^p(\rr^d)$ is the usual Lebesgue space with
$\llVert f\rrVert _{L^p(\rr^d)}:=  (\int_{\rr^d}\llvert f(x)\rrvert ^p\,dx  )^{1/p}$.

\item For $s\in(0,1)$, the Besov space $B^s_{1,\infty}(\rr^d)$
consists of all
functions $f$ such that
\[
\llVert f\rrVert _{B^s_{1,\infty}(\rr^d)}:=\llVert f\rrVert _{L^1(\rr^d)}+ \sup
_{h\in\rr^d,
0<\llvert h\rrvert <1} \llvert h\rrvert ^{-s} \int
_{\rr^d}\bigl\llvert f(x+h)-f(x)\bigr\rrvert \,dx<\infty.
\]
\end{itemize}

In the whole paper, when a measure $f\in\cM(\rr^d)$ has a density, we
also denote by $f$ this density.

\subsection{Weak solutions}

We will consider weak solutions in the following sense.

\begin{defi}\label{dfws}
Assume (\ref{aaa}) for some $\nu\in(0,1)$ and $\gamma\in(-1,1)$.

(i) A family $(f_t)_{t\geq0} \subset\cP_2(\rr^3)$
is a weak solution to (\ref{be}) if for all $t\geq0$,
\renewcommand\theequation{\thesection.\arabic{equation}}
%
\begin{eqnarray}
\label{energy} \hspace*{25pt}\int_{\rr^3}v f_t(dv)=
\int_{\rr^3}v f_0(dv) \quad\mbox{and}\quad \int
_{\rr^3}\llvert v\rrvert ^2 f_t(dv)=
\int_{\rr^3}\llvert v\rrvert ^2 f_0(dv)<
\infty
\end{eqnarray}
and if for any $\phi\in\lip_b(\rr^3)$ and any $t\geq0$,
%
\begin{eqnarray}
\label{wbe} &&\int_{\rr^3}\phi(v)f_t(dv)
\nonumber
\\[-8pt]
\\[-8pt]
&&\qquad= \int_{\rr^3}\phi(v)f_0(dv) + \int
_0^t \int_{\rr^3}\int
_{\rr^3}L_B \phi(v,v_*) f_s(dv_*)f_s(dv)
\,ds ,
\nonumber
\end{eqnarray}
where, for $v'=v'(v,v_*,\sigma)$ and $\theta=\theta(v,v_*,\sigma)$
defined in (\ref{vprime}),
%
\begin{eqnarray}
\label{dfL} L_B\phi(v,v_*):= \int_{{\mathbb{S}^2}} B
\bigl(\llvert v-v_*\rrvert ,\cos\theta\bigr) \bigl[\phi \bigl(v'
\bigr)-\phi (v) \bigr] \,d\sigma.
\end{eqnarray}
\end{defi}

The right-hand side of (\ref{wbe}) is well-defined due to (\ref
{energy}) and (\ref{aaa}).
Indeed, there holds $\llvert v'-v\rrvert =\llvert v-v_*\rrvert \sqrt{(1-\cos\theta)/2} \leq\llvert v-v_*\rrvert
\llvert \theta\rrvert $,
so that $\llvert L_B\phi(v,\break v_*)\rrvert \leq C_\phi\int_{{\mathbb{S}^2}}
B(\llvert v-v_*\rrvert ,\vspace*{1pt}\cos
\theta)
\llvert v-v_*\rrvert  \llvert \theta\rrvert  \,d\sigma
\leq C_\phi\llvert v-v_*\rrvert ^{1+\gamma} \int_0^{\pi/2} \llvert \theta\rrvert ^{-\nu
}\,d\theta
\leq C_\phi(1+\llvert v\rrvert ^2+\llvert v_*\rrvert ^2)$.

Concerning the well-posedness of (\ref{be}) given $f_0\in\cP_2(\rr^3)$,
the following results are available.

\subsubsection*{Hard potentials} Assume (\ref{aaa}) for some $\nu\in(0,1)$
and $\gamma\in(0,1)$.
Then by Lu--Mouhot \cite{lm}, there exists a weak solution to (\ref
{be}) starting from
$f_0$. This solution furthermore satisfies that $\sup_{[t_0,\infty)}
m_p(f_t) <\infty$
for all $t_0>0$, all $p\geq2$. Such a moment production property was
discovered by Elmroth \cite{e}
and Desvillettes \cite{d}. Two different uniqueness results are available,
assuming either that $f_0$ is regular ($f_0\in W^{1,1}(\rr^3)$ with
$\int_{\rr^3}(1+\llvert v\rrvert ^2) \llvert \nabla f_0(v)\rrvert
\,dv<\infty$,
Desvillettes and Mouhot \cite{dm}) or localized ($\int_{\rr^3}e^{a
\llvert v\rrvert ^\gamma
}f_0(dv)<\infty$ for some $a>0$,
\cite{fm}).

\subsubsection*{Maxwell molecules} Assume (\ref{aaa}) for some $\nu\in
(0,1)$ and with $\gamma=0$.
Then there exists a unique weak solution to (\ref{be}) starting from
$f_0$ due to Toscani and Villani \cite{tv}.

\subsubsection*{Moderately soft potentials} Assume (\ref{aaa}) for some
$\nu\in(0,1)$, some
$\gamma\in(-1,0)$ with $\gamma+\nu>0$. Assume also that $f_0$ has a
density with a finite
entropy, that is, $\int_{\rr^3}f_0(v)\llvert \log f_0(v)\rrvert \,dv <\infty$.
Then there exists a weak solution to (\ref{be}) starting from $f_0$ due
to Villani \cite{v:nc}.
This solution is unique \cite{fm} if $f_0 \in\cP_q(\rr^3)$ for some
$q> \gamma^2/(\gamma+\nu)$.

\subsubsection*{Very soft potentials} Assume (\ref{aaa}) for some $\nu
\in
(0,2)$, some
$\gamma\in(-3,0)$. If $f_0$ has a density with a finite
entropy, there exists a weak solution to (\ref{be}) starting from $f_0$
due to Villani \cite{v:nc}.
Uniqueness holds locally in time \cite{fgu} provided $f_0 \in L^p(\rr
^3)$ for some $p>3/(3+\gamma)$.

\subsection{Main result}
Let us mention that during the proof, we will
check the following property.

\begin{theorem}\label{support}
Assume (\ref{aaa}) for some $\gamma\in(-1,1)$,
$\nu\in(0,1)$. Let also $f_0 \in\cP_2(\rr^3)$ not be a Dirac mass.
For any weak solution $(f_t)_{t\geq0}$ to (\ref{be}) starting from $f_0$,
$\supp f_t = \rr^3$ for all $t>0$.
\end{theorem}

The main result of the paper is the following.

\begin{theorem}\label{mr}
Assume (\ref{aaa}) for some $\gamma\in(-1,1)$, $\nu\in(0,1)$
with $\gamma+\allowbreak\nu>0$.
Let $f_0 \in\cP_2(\rr^3)$ not be a Dirac mass.
\begin{longlist}[(iii)]
\item[(i)] If $\gamma\in(0,1)$, then any weak solution $(f_t)_{t\geq0}$ to
(\ref{be}) starting from
$f_0$ and such that
%
\begin{eqnarray}
\label{mom} \forall t_0>0 , \forall p \geq2, \qquad\sup
_{t\geq t_0} m_p(f_t)<\infty
\end{eqnarray}
satisfies that $f_t \in B^s_{1,\infty}(\rr^3)$ for all $t>0$, all
$s \in(0,s_\nu)$,
where
%
\begin{eqnarray}
\label{snu} s_\nu&=& \sup_{\alpha\in(0,\nu]} \biggl(
\frac{2\alpha}{1+2\alpha
}-\alpha \biggr)
\nonumber
\\[-8pt]
\\[-8pt]
&=& \cases{ \displaystyle\bigl(\nu-2\nu^2\bigr)/(1+2\nu) &\quad if $
\nu\in\bigl(0,(\sqrt2-1)/2\bigr)$,
\cr
\displaystyle(\sqrt2 - 1)^2/2 &
\quad if $\nu\in\bigl[(\sqrt2-1)/2,1\bigr)$.}
\nonumber
\end{eqnarray}

\item[(ii)] If $\gamma\in(-1,0]$, assume also that $f_0 \in\cP_{4+\gamma
+4\llvert \gamma\rrvert /\nu}(\rr^3)$.
There exists a weak solution $(f_t)_{t\geq0}$ to (\ref{be}) starting from
$f_0$ such that $f_t \in B^s_{1,\infty}(\rr^3)$ for all $t>0$, all
$s \in(0,s_{\gamma,\nu})$,
where
%
\begin{eqnarray}
\label{sganu} s_{\gamma,\nu} = \sup_{\alpha\in(0,\nu]} \biggl(
\frac{(2+\gamma
/\nu
)\alpha}{1+(2+\gamma/\nu)\alpha}-\alpha \biggr).
\end{eqnarray}

\item[(iii)] In any case, $f_t$ has a density satisfying $\int_{\rr^3}f_t(v)
\llvert \log
f_t(v)\rrvert \,dv <\infty$ as soon as $t>0$.
\end{longlist}
\end{theorem}

No regularization may hold if $f_0$ is a Dirac mass, since Dirac masses
are stationary solutions to
(\ref{be}).
In the case of moderately soft potentials ($\gamma\in(-1,0]$ and
$\gamma+\nu>0$),
we need a few moments; observe that we always
have $4 \leq4+\gamma+4\llvert \gamma\rrvert /\nu\leq8$.
Of course, (\ref{sganu}) can of be made explicit, but the resulting
formula is awful.
While we show that any solution is regularized for hard potentials,
we can only prove that there exists
at least one solution enjoying some regularization properties for
moderately soft
potentials. This is due to our probabilistic
interpretation: when $\gamma\in(0,1)$, we can associate a Boltzmann
stochastic process to any weak
solution, while when $\gamma\in(-1,0]$, we are only able to prove
that there exists a
Boltzmann stochastic process and that its law is a weak solution.

In \cite{v:h}, Theorem~9(iii), page 95,  Villani announces a result
very similar to Theorem~\ref{mr}.
However, he obtains only some gain of integrability, while we obtain some
(extremely weak) regularity.
We know from a private communication that this work has never been
written down.

\begin{rk} As can be checked from the proof, the same result as
stated in Theorem~\ref{mr}(i)
holds for regularized hard potentials where $B(\llvert v-v_*\rrvert ,\cos\theta
)=(1+\llvert v-v_*\rrvert ^2)^{\gamma/2}b(\theta)$,
with $\gamma\in(0,1)$ and $c_0 \llvert \theta\rrvert ^{-\nu-1} \leq b(\theta)
\leq
C_0 \llvert \theta\rrvert ^{-\nu-1}$ for some
$\nu\in(0,1)$.
\end{rk}

\subsection{Motivation}
The main interest of Theorem~\ref{mr} is the following: almost all the
papers on the Boltzmann equation
(concerning, e.g., regularization or large-time behavior) assume that
the initial condition
has a finite entropy; see the long review paper of Villani \cite{v:h}.
This condition is of
course physically reasonnable. Our result shows that it is indeed
physically reasonnable,
since the entropy \textit{automatically} becomes finite.
Consequently, the results assuming the finiteness of the entropy of the
initial condition
extend to any measure initial
data with a finite mass and energy which are not Dirac masses. For example,
we deduce from Alexandre et al.
\cite{advw}, Chen and He \cite{ch}, Desvillettes and Wennberg \cite{dw}
and Huo et al. \cite{hmuy}
that for any (non-Dirac) measure initial condition
with finite mass and energy:
\begin{itemize}
\item under the assumptions of Theorem~\ref{mr},
$(1+\llvert v\rrvert ^2)^{\gamma/2}\sqrt{f_t(v)} \in H^{\nu/2}(\rr^3)$ for all $t>0$
by \cite{ch};

\item for regularized hard potentials, $f_t \in C^\infty(\rr^3)$
for all $t>0$ due to \cite{dw,hmuy}.
\end{itemize}

\subsection{Known regularization results}

In many papers, Grad's cutoff is assumed: the cross section $B$, which
physically satisfies
$\int_0^{\pi} B(\llvert v-v_*\rrvert ,\cos\theta)\,d\theta= \infty$,
is replaced by an integrable cross section.
No regularization may arise under Grad's cutoff; see, for example,
Mouhot and Villani \cite{mv}.
The first results about regularization for the homogeneous Boltzmann
equation without cutoff
are due to Desvillettes \cite{d1,d2}.
There are now roughly four types of available results.
\begin{itemize}
\item General results applying to all \textit{true} physical
potentials, relying on the entropy dissipation,
providing weak regularity.
Under (\ref{aaa}) for some $\nu\in(0,2)$ and some $\gamma
\in
(-3,1)$, when $f_0$
is a function with finite mass, entropy and energy,
it has been shown (among many other things) by Alexandre et al.
\cite{advw} that $\sqrt{f_t} \in H^{\nu/2}_{\mathrm{loc}}(\rr^3)$ for all $t>0$.
This has been recently precised,
in the case of hard and moderately soft potentials by
Chen and He \cite{ch}, Theorem~1.3: $(1+\llvert v\rrvert ^2)^{\gamma/2}\sqrt{f_t(v)}
\in H^{\nu/2}(\rr^3)$ for all $t>0$.

\item High regularization for \textit{true} physical potentials assuming
that $f$ is already known to be slightly regular.
It is proved by Chen and He \cite{ch}, Theorem~1.5, that for hard and
moderately soft potentials,
if $f_0 \in H^3(\rr^3)$ and $\int_{\rr^3}(1+\llvert v\rrvert ^q) \llvert \nabla
f_0(v)\rrvert \,dv<\infty$
for some $q\geq2$
large enough, then the solution immediately lies in $H^N(\rr^3)$ for
some $N$ depending on $q$.

\item Full regularization for \textit{regularized} hard potentials,
when $f_0$
is a function with finite mass, entropy and energy. See Desvillettes
and Wennberg \cite{dw},
Alexandre and Elsafadi \cite{aes2} and Huo et al. \cite{hmuy}.

\item Very restrictive results when $f_0$ is a (non-Dirac)
probability measure in the $2D$ case:
full regularization for Maxwell molecules (see Graham and M\'el\'eard
\cite{gm} and
\cite{f:r2d}) and weak regularization \cite{bf} for a class of hard
potentials (applying to interaction forces in $1/r^s$ with $s>13.75$).
All these works use some Malliavin calculus and seem very difficult
to extend to the $3D$ case.
\end{itemize}

Here we deal with \textit{true} physical potentials, for which there are
several complications:
$\llvert w\rrvert ^\gamma$ is not bounded below (and vanishes when $\gamma>0$), which
makes ellipticity estimates
nontrivial, explodes either at $0$ or at infinity and is in any case
not smooth at $0$.
To our knowledge, the only regularization results that concern
the homogeneous Boltzmann equation for
true physical potentials are those of
\cite{advw}, \cite{ch} and \cite{bf}.
The present result consequently improves on \cite{bf} (we treat the
$3D$ case, all
interaction forces in $1/r^s$ with $s>3$ and we remove some technical
assumptions)
and is not in competition
with \cite{advw} or \cite{ch} (the finiteness of the entropy is assumed
in \cite{advw}
and \cite{ch}).

\subsection{Known positivity results}

The proof of Theorem~\ref{support} is very easy, but it seems to be new.
The first lower bound of solutions to the Boltzmann equation is due to
Carleman \cite{ca} in the case of
hard spheres ($\gamma=1$, $b \equiv1$).
In \cite{pw}, Pulvirenti and Wennberg obtained some Maxwellian
lowerbound in the case of hard potentials with cutoff
($\gamma\in(0,1]$ and $\int_0^\pi b(\theta)\,d\theta<\infty$), assuming
that $f_0$ has a finite entropy.
A quantitative version of Theorem~\ref{support} (for measure solutions)
has been proved by Zhang and Zhang \cite{zz}, still
in the case of hard potentials with cutoff.
Some positivity results \cite{f:p} are available for $2D$ Maxwell
molecules without cutoff.
For general physical potentials without cutoff, some indications
concerning the positivity
of smooth solutions are given in Villani \cite{v:h}, Sections~6.2 and 6.3.
Finally, Mouhot \cite{m} proved some quantitative lower bound in the
much more complicated
spatially inhomogeneous
case without cutoff, but for quite regular solutions [corresponding
here, roughly, to the assumption
$f \in L^\infty_{\mathrm{loc}}([0,\infty),W^{2,\infty}(\rr^3))$].

\subsection{Comments on the method}

The classical way to prove some regularization results by probabilistic
methods is to
use some Malliavin calculus, based on the famous probabilistic
interpretation of
the homogeneous Boltzmann equation in terms of a nonlinear jumping
stochastic differential equation
initiated by Tanaka \cite{t}.
Unfortunately, this s.d.e. has regular coefficients only in the $2D$-case
and for Maxwell molecules. In the case of $3D$ Maxwell molecules, a
sort of Lipschitz property was observed by Tanaka \cite{t}
(see Lemma~\ref{tanana} below), but we cannot hope for more.
This seems to make almost impossible the use of Malliavin calculus to
study the $3D$ Boltzmann equation.

Here we use no Malliavin calculus, but a recent method
introduced in \cite{fp} to prove that stochastic processes with rather
irregular coefficients
have a density.
Recently, Debussche and Romito \cite{dr} have considerably improved
this method
by using Besov spaces, in order to study the regularity of the
law of the solution to a $3D$ stochastic Navier--Stokes equation.
For example, only $1D$ diffusion processes with diffusion coefficient in
$C^{1/2+\epsilon}_b(\rr)$
were treated in \cite{fp}, while some quick computations seem to show that
diffusion processes in any dimension and with
diffusion coefficient in $C^{\epsilon}_b(\rr^d)$ can be studied using the
tools of \cite{dr}.
As we will see, it also perfectly applies to the s.d.e. associated with
the homogeneous Boltzmann
equation.

Let us mention that our proof is not \textit{deeply} probabilistic:
we use no stopping times,
no Malliavin calculus, etc. We believe that a very similar
deterministic proof can be written down. The advantage would be to
remove Section~\ref{exifcs}
below, which is long and boring, in which we build the stochastic
processes related to Boltzmann's equation.
The disadvantage would be that the computations of Section~\ref{appro}
would become awful (and
would look completely artificial).

\subsection{Heuristics}

Let us say a word about the reasons for regularization. Consider an
initial velocity
distribution $f_0$, possibly very singular.
Pick at random a particle in the initial system, and call $V_t$ its
velocity at time $t$. Observe that
the law of $V_t$ is $f_t$ for all $t\geq0$.
This particle collides, at time $t\geq0$, at rate $\int_{\rr^3}\int_{{\mathbb{S}^2}}
B(\llvert V_t-v_*\rrvert ,\cos\theta)\,d\sigma f_t(dv_*)$.
In the case without cutoff, this rate is thus infinite: the particle is
subjected to infinitely many
collisions on each finite time interval. Furthermore, at each
collision, some randomness is added, since
$v_*$ and $\sigma$ are chosen at random. Hence, we expect that for each
$t>0$, our particle has been subjected
to infinitely many collisions on the time interval $[0,t]$, each of
these collisions producing some randomness.
Consequently, $V_t$ will be much more random than $V_0$, so that its
law should be much more regular.

Conversely, in the case with cutoff where the rate of collision of our
particle is finite,
we expect that $V_t=V_0$ during some (random) positive time, so that
the solution $f_t$ will contain all
the singularities of $f_0$, at least for small times.

\subsection{Plan of the paper}

In the next section, we state the main lemma we will use, which is due
to Debussche and Romito \cite{dr}
and we give an elementary proof.
In Section~\ref{wsp}, we rewrite in an adequate way the weak
formulation of (\ref{be})
and prove a few properties of weak solutions.
Section~\ref{supp} is devoted to the proof of Theorem~\ref{support} and
to some slightly
more quantitative lower bound.
Then we adapt the probabilistic interpretation of Tanaka \cite{t} to
hard and
moderately soft potentials in Section~\ref{proba}.
The proof of the existence of the Boltzmann process
lies at the end of the paper (Section~\ref{exifcs}).
Then the strategy of the proof is the following: we approximate the
Boltzmann process
by a L\'evy process (Section~\ref{appro}) and
study the regularity of the law of the approximating L\'evy process
(Section~\ref{reglev}).
Using that the approximating process has a regular law and that the
true Boltzmann process
is close to the approximating process, we conclude in Section~\ref{conclu}.

\subsection{Notation}

We will write $C$ for a (large) finite constant and $c$ for a (small)
positive constant, whose
values may change from line to line and which depend only on $\nu
,\gamma
,c_0,C_0$ [recall
(\ref{aaa})] and on the weak solution $(f_t)_{t\geq0}$.
We write in index all the additional dependence of constants.

\section{Main lemma}\label{ml}

Our study is based on the following result due to Debussche and Romito
\cite{dr}, End of the proof of Theorem~5.1.

\begin{lem}\label{thelem}
Let $g \in\cM(\rr^d)$. Assume that there are $0<\alpha<a<1$ and a
constant $\kappa$ such that for all
function $\phi\in C^\alpha_b(\rr^d)$, all $h \in\rr^d$ with
$\llvert h\rrvert \leq1$,
%
\begin{eqnarray}
\label{tla} \biggl\llvert \int_{\rr^d}\bigl[\phi(x+h)-
\phi(x)\bigr] g(dx) \biggr\rrvert \leq\kappa \llVert \phi \rrVert _{C^\alpha_b(\rr^d)}
\llvert h \rrvert ^a.
\end{eqnarray}
Then $g$ has a density in $B^{a-\alpha}_{1,\infty}(\rr^d)$ and
$ \llVert g\rrVert _{B^{a-\alpha}_{1,\infty}(\rr^d)} \leq g(\rr^d) +
C_{d,a,\alpha}
\kappa$.
\end{lem}

Actually, the result in \cite{dr} is more general. The proof in \cite
{dr} relies
on several theorems of functional analysis. We present here an \textit{elementary}
(though longer) proof.

\begin{pf*}{Proof of Lemma~\ref{thelem}}
We divide the proof into four steps.

\textit{Step 1: Preliminaries.} For $r>0$,
consider the function $\chi_r(x)=(v_d r^d)^{-1}\*\indiq_{\{\llvert x\rrvert <r\}}$,
where $v_d$ is the volume of the unit ball in $\rr^d$. An easy
computation shows that for all
$x, y \in\rr^d$,
%
\begin{eqnarray}
\label{ll1} \int_{\rr^d}\bigl\llvert \chi_r(x-z)-
\chi_r(y-z)\bigr\rrvert \,dz \leq C_d \min\bigl( 1,
\llvert x-y\rrvert /r\bigr).
\end{eqnarray}
For $\psi\in L^\infty(\rr^d)$ and $r\in(0,1]$, $\psi\star\chi_r$
belongs to $C^\alpha_b(\rr^d)$
(it is actually
Lipschitz-continuous) and
%
\begin{eqnarray}
\label{ll2} \llVert \psi\star\chi_r \rrVert _{C^\alpha_b(\rr^d)}
\leq C_d \llVert \psi \rrVert _{L^\infty
(\rr^d)} r^{-\alpha}.
\end{eqnarray}
Indeed, it obviously holds that $\llVert \psi\star\chi_r \rrVert _{L^\infty(\rr
^d)}\leq\llVert \psi\rrVert _{L^\infty(\rr^d)}$
and for $x \ne y$, we deduce from (\ref{ll1}) that
$\llvert \psi\star\chi_r (x) - \psi\star\chi_r (y)\rrvert \leq C_d
\llVert \psi\rrVert _{L^\infty(\rr^d)} \min( 1, \llvert x-y\rrvert /r) \leq C_d \llVert \psi
\rrVert _{L^\infty(\rr^d)}r^{-\alpha} \llvert x-y\rrvert ^{\alpha}$.

\textit{Step 2.} Next we prove that for any $r\in(0,1]$,
any $\llvert h\rrvert \leq1$,
\[
\int_{\rr^d}\bigl\llvert g\star\chi_r (x+h)- g
\star\chi_r (x)\bigr\rrvert \,dx \leq C_d \kappa \llvert
h\rrvert ^a r^{-\alpha}. %
\]
It suffices to prove that for any $\psi\in L^\infty(\rr^d)$,
$I_r(h,\psi):=\llvert \int_{\rr^d}\psi(x)[g\star\chi_r (x+h)- g\star
\chi_r (x)]\,dx \rrvert
\leq C_d \kappa\llVert \psi\rrVert _{L^\infty(\rr^d)}\llvert h\rrvert ^a r^{-\alpha}$. But using
(\ref{tla}) and (\ref{ll2}), we get
\begin{eqnarray*}
I_r(h,\psi)&=&\biggl\llvert \int_{\rr^d}\bigl[\psi
\star\chi_r(y-h)- \psi\star \chi _r(y) \bigr] g(dy)
\biggr\rrvert \leq\kappa\llVert \psi\star\chi_r \rrVert
_{C^\alpha_b(\rr^d)} \llvert h\rrvert ^a \\
&\leq& C_d \kappa
\llVert \psi\rrVert _{L^\infty(\rr^d)} \llvert h\rrvert ^a
r^{-\alpha}. %
\end{eqnarray*}

\textit{Step 3.} Here we assume additionally that $g$ has a density in
$C^1(\rr^d)$ satisfying
$\int_{\rr^d}\llvert \nabla g(x)\rrvert \,dx <\infty$ (which implies that all the
computations below are licit),
and we check that
\[
\sup_{\llvert h\rrvert \leq1}\llvert h\rrvert ^{\alpha-a}\int
_{\rr^d}\bigl\llvert g(x+h)-g(x)\bigr\rrvert \,dx \leq
C_{d,a,\alpha
}\kappa. %
\]
To this end, we first write, using Step 2, for all $\llvert h\rrvert \leq1$, all
$r\in(0,1]$,
\begin{eqnarray*}
&&\int_{\rr^d}\bigl\llvert g(x+h)-g(x)\bigr\rrvert \,dx\\
&&\qquad\leq
\int_{\rr^d}\bigl\llvert g\star\chi_r (x+h)- g
\star\chi_r (x)\bigr\rrvert \,dx + 2 \int_{\rr^d}
\bigl\llvert g\star\chi_r (x)- g(x)\bigr\rrvert \,dx
\\
&&\qquad\leq C_d \kappa\llvert h\rrvert ^a r^{-\alpha} +
\frac{2}{v_d r^d} \int_{\rr
^d}\int_{\rr^d}
\bigl\llvert g(y)-g(x)\bigr\rrvert \indiq_{\{\llvert x-y\rrvert <r\}}\,dx\,dy
\\
&&\qquad=C_d \kappa\llvert h\rrvert ^a r^{-\alpha} +
\frac{2}{v_d r^d} \int_{\llvert u\rrvert <r} \,du \int_{\rr^d}dx
\bigl\llvert g(x+u)-g(x)\bigr\rrvert .
\end{eqnarray*}
Thus, setting $I_t:=\sup_{\llvert h\rrvert =t} \int_{\rr^d}\llvert g(x+h)-g(x)\rrvert \,dx$ and
$S_t=\sup_{s\in(0,t]}s^{\alpha-a}I_s$,
we deduce that for all $t\in(0,1]$, all $r\in(0,1]$ (below, the
variable $u$ belongs to $\rr^d$),
\begin{eqnarray*}
t^{\alpha-a}I_t&\leq& C_d \kappa(t/r)^{\alpha}
+ \frac{2t^{\alpha
-a}}{v_d r^d} \int_{\llvert u\rrvert <r} \llvert u\rrvert
^{a-\alpha} S_{\llvert u\rrvert }\,du
\\
&\leq& C_d \kappa(t/r)^{\alpha}+ \frac{2t^{\alpha-a}}{v_d r^d}
S_1 r^{a-\alpha} v_d r^d
\\
&\leq& C_d \kappa(t/r)^{\alpha}+ 2 (r/t)^{a-\alpha}
S_1.
\end{eqnarray*}
Choosing $r=4^{-1/(a-\alpha)} t$, we deduce that
for all $t\in(0,1]$, $t^{\alpha-a}I_t \leq4^{\alpha/(a-\alpha)}\*C_d
\kappa+ S_1/2$.
This implies $S_1 \leq4^{\alpha/(a-\alpha)}C_d \kappa+ S_1/2$ and finally
$S_1 \leq\break 2.4^{\alpha/(a-\alpha)}\*C_d \kappa$ as desired.

\textit{Step 4.} Consider now $g$ as in the statement. For $n\geq1$, put
$g_n=g\star G_n$, where
$G_n(x)=(n/\pi)^{d/2} e^{-n\llvert x\rrvert ^2}$. Then $g_n \in C^1(\rr^d)$,
$\int_{\rr^d}g_n(x)\,dx=g(\rr^d)$ and $\int_{\rr^d}\llvert \nabla g_n(x)\rrvert \,dx
<\infty$.
Furthermore, one easily checks
that $g_n$ satisfies (\ref{tla}) with the same constant $\kappa$ as $g$.
Thus we can apply Step 3 and deduce that
$\sup_{\llvert h\rrvert \leq1}\llvert h\rrvert ^{\alpha-a}\int_{\rr^d}\llvert g_n(x+h)-g_n(x)\rrvert \,dx \leq
C_{d,a,\alpha}\kappa$ for all $n\geq1$,
whence $\llVert g_n\rrVert _{B^{a-\alpha}_{1,\infty}} \leq g(\rr^d)+
C_{d,a,\alpha
}\kappa$ (recall Section~\ref{fs}).
Consequently, the sequence
$g_n$ is strongly compact in $L^1(\rr^d)$ (because the balls of
$B^s_{1,\infty}(\rr^d)$ are compact in
$L^1(\rr^d)$ for all $s>0$; see, e.g., \cite{rs}). But $g_n$ tends
weakly (in the sense of measures)
to $g$. We deduce that $g \in L^1(\rr^d)$ and that we can find a
subsequence such that
$\lim_k \llVert g_{n_k}-g\rrVert _{L^1(\rr^d)}=0$. One easily concludes that for all
$\llvert h\rrvert \leq1$, $\int_{\rr^d}\llvert g(x+h)-g(x)\rrvert \,dx=\lim_k \int_{\rr^d}
\llvert g_{n_k}(x+h)-g_{n_k}(x)\rrvert \,dx \leq C_{d,a,\alpha}\kappa
\llvert h\rrvert ^{a-\alpha}$.\vspace*{2pt} We deduce that $\llVert g\rrVert _{B^{a-\alpha}_{1,\infty}(\rr^d)}
\leq g(\rr^d)+ C_{d,a,\alpha}\kappa$.
\end{pf*}

\section{Weak solutions}\label{wsp}

First, we parameterize (\ref{vprime}) as in \cite{fsm}.
For each $X\in\rr^3\setminus\{0\}$, we introduce $I(X),J(X)\in\rr^3$
such that
$(\frac{X}{\llvert X\rrvert },\frac{I(X)}{\llvert X\rrvert },\frac{J(X)}{\llvert X\rrvert })$
is an orthonormal basis of $\rr^3$, in such a way that $X\mapsto
(I(X),J(X))$ is measurable.
We also put $I(0)=J(0)=0$.
For $X,v,v_*\in\rr^3$, $\theta\in[0,\pi)$ and $\varphi\in[0,2\pi)$,
we set
%
\begin{eqnarray}
\label{dfvprime} \cases{ \displaystyle\Gamma(X,\varphi):=(\cos
\varphi) I(X) + (\sin\varphi)J(X),
\cr\displaystyle
v'(v,v_*,\theta,\varphi):= v - \frac{1-\cos\theta}{2} (v-v_*) +
\frac{\sin\theta}{2}\Gamma(v-v_*,\varphi),
\cr\displaystyle
a(v,v_*,\theta,\varphi):=v'(v,v_*,\theta,\varphi)-v. }
\end{eqnarray}
The choice of $(I(X),J(X))$ does not matter. The important thing is that
for any reasonable $F\dvtx \rr^3\times\rr^3\times\rr^3\times[0,\pi
)\mapsto
\rr$, any
$v,v_* \in\rr^3$,
\[
\int_0^\pi\int_0^{2\pi}
F\bigl(v,v_*,v'(v,v_*,\theta,\varphi),\theta\bigr) \sin \theta \,d
\varphi \,d\theta = \int_{{\mathbb{S}^2}} F\bigl(v,v_*,v',
\theta\bigr)\,d\sigma,
\]
where on the right-hand side, $v'=v'(v,v_*,\sigma)$ and
$\theta=\theta(v,v_*,\sigma)\in(0,\pi)$ are defined by
(\ref{vprime}). This in particular implies that for all $\phi\in\lip
_b(\rr^3)$, recalling (\ref{dfL})
and then (\ref{aaa}),
%
\begin{eqnarray}\label{newL}
&&\hspace*{20pt}L_B\phi(v,v_*)\nonumber\\[-8pt]\\[-8pt]
&&\hspace*{20pt}\qquad= \int_0^\pi
\int_0^{2\pi} \bigl[\phi\bigl(v+a(v,v_*,\theta ,
\varphi )\bigr) - \phi(v)\bigr] B\bigl(\llvert v-v_*\rrvert ,\cos\theta\bigr)\sin
\theta \,d\varphi \,d\theta\nonumber
\\\label{newL2}
&&\hspace*{20pt}\qquad= \llvert v-v_*\rrvert ^\gamma\int_0^{\pi/2}
\int_0^{2\pi} \bigl[\phi \bigl(v+a(v,v_*,\theta ,
\varphi)\bigr) - \phi(v)\bigr] b(\theta)\,d\varphi\, d\theta.
\end{eqnarray}
We will frequently use that, by a straightforward computation,
%
\begin{eqnarray}
\label{lit} \bigl\llvert a(v,v_*,\theta,\varphi)\bigr\rrvert = \sqrt{
\frac{1-\cos\theta}{2}} \llvert v-v_*\rrvert \leq \frac{1} 2 \theta\llvert
v-v_*\rrvert .
\end{eqnarray}
We will also need the following remark, corresponding to the $2D$ equality
$ \langle\xi, X^\perp \rangle= \pm \langle\xi
^\perp, X  \rangle$.

\begin{rk}\label{perp}
For any measurable nonnegative function $F\dvtx  \rr\mapsto\rr$, any
$X\in
\rr^3$,
any $\xi\in\rr^3$,
\[
\int_0^{2\pi} F\bigl( \bigl\langle\xi,\Gamma(X,
\varphi) \bigr\rangle \bigr)\,d\varphi= \int_0^{2\pi}
F\bigl( \bigl\langle X,\Gamma(\xi,\varphi) \bigr\rangle\bigr)\,d\varphi.
\]
\end{rk}

\begin{pf}
Recall that these integrals do not depend on the choice of
$(I(X),\allowbreak J(X))$ and $(I(\xi), J(\xi))$ [as soon as $(\frac
{X}{\llvert X\rrvert },\frac
{I(X)}{\llvert X\rrvert },\frac{J(X)}{\llvert X\rrvert })$
and $(\frac{\xi}{\llvert \xi\rrvert },\frac{I(\xi)}{\llvert \xi\rrvert },\frac{J(\xi)}{\llvert \xi
\rrvert })$ are
orthonormal bases of $\rr^3$].
If $X$ and $\xi$ are colinear
$ \langle\xi,\Gamma(X,\varphi) \rangle=  \langle
X,\Gamma(\xi,\varphi) \rangle=0$
for all
$\varphi$ and the result follows.
Otherwise, choose $(I(X),J(X))$ and $(I(\xi),J(\xi))$ such
that $X,\xi,I(X),I(\xi)$ are in the same plane and such that $
\langle
X,I(\xi) \rangle= \langle\xi,I(X) \rangle$,
which implies that
$ \langle\xi,\Gamma(X,\varphi) \rangle=  \langle
X,\Gamma(\xi,\varphi) \rangle$
for all
$\varphi$.
\end{pf}

Unfortunately, it is not possible to build $I$ in such a way that
$X\mapsto I(X)$ is smooth.
Tanaka \cite{t} found a way to overcome this difficulty, which was
slightly precised in \cite{fsm}, Lemma~2.6.

\begin{lem}\label{tanana}
There exists a measurable function $\varphi_0 \dvtx  \rr^3 \times\rr^3
\mapsto[0,2\pi)$, such that for all
$v,v_*,w,w_* \in\rr^3$, all $\theta\in[0,\pi)$ and all
$\varphi\in[0,2\pi)$,
\[
\bigl\llvert a(v,v_*,\theta,\varphi) - a\bigl(w,w_*,\theta,\varphi+
\varphi_0(v-v_*,w-w_*)\bigr)\bigr\rrvert \leq2 \theta \bigl( \llvert
v-w\rrvert + \llvert v_* - w_*\rrvert \bigr).
\]
\end{lem}

We conclude this section with a useful time-regularity property of weak
solutions. This must be more or
less classical; see, for example, Gamba, Panferov and Villani \cite
{gpv} for a stronger result in the case of
cutoff hard potentials, but we found no precise reference in the
present setting.

\begin{lem}\label{trws}
Let $f_0 \in\cP_2(\rr^3)$. Assume (\ref{aaa}) for some
$\gamma
\in(-1,1)$,
$\nu\in(0,1)$. Consider any weak solution
$(f_t)_{t\geq0}$ to (\ref{be}) starting from $f_0$. Then for any
$\phi\in\lip_b(\rr^3)$, $L_B\phi$ is continuous on $\rr^3\times
\rr^3$ and
the map $t \mapsto\int_{\rr^3}\phi(v) f_t(dv)$ belongs to
$C^1([0,\infty))$.
\end{lem}

\begin{pf}
Recall (\ref{wbe}): to show that $t \mapsto\int_{\rr^3}\phi(v)
f_t(dv)$ is
of class $C^1([0,\infty))$,
it suffices to check that
$t \mapsto\int_{\rr^3}\int_{\rr^3}L_B\phi(v,v_*)
f_t(dv_*)f_t(dv)$ is continuous
on $[0,\infty)$.

\textit{Step 1.} For $\phi\in\lip_b(\rr^3)$, $\llvert L_B\phi(v,v_*)\rrvert  \leq
C_\phi
\llvert v-v_*\rrvert ^{\gamma+1}\leq
C_\phi(1+\llvert v\rrvert ^2+\llvert v_*\rrvert ^2)$ by (\ref{newL2}), (\ref{lit}) and since
$\int_0^{\pi/2} \theta
b(\theta)\,d\theta<\infty$ by (\ref{aaa}).
By (\ref{energy}), we deduce that $\int_{\rr^3}\int_{\rr^3}L_B\phi
(v,v_*)f_t(dv_*)f_t(dv)$ is bounded, so that
$t \mapsto\int_{\rr^3}\phi(v)\*f_t(dv)$ is continuous on $[0,\infty
)$ by
(\ref{wbe}).
The Portemanteau theorem thus implies that $t \mapsto f_t$ is weakly
continuous, which
classically implies that $t \mapsto f_t\otimes f_t$ is weakly continuous:
for all $\phi\in C_b(\rr^3\times\rr^3)$, $t\mapsto\int_{\rr
^3}\int_{\rr^3}
\phi(v,v_*)f_t(dv)\*f_t(dv_*)$ is continuous on $[0,\infty)$.

\textit{Step 2.} Recall that $B(\llvert v-v_*\rrvert ,\cos\theta)\sin\theta
=\llvert v-v_*\rrvert ^\gamma b(\theta)$ by (\ref{aaa}) and
define, for $k \geq1$, $B_k(\llvert v-v_*\rrvert ,\cos\theta)\sin\theta
=(\llvert v-v_*\rrvert ^\gamma\land k) b(\theta)
\indiq_{\{\theta>1/k\}}$.
It is immediately checked that $L_{B_k}\phi\in C_b(\rr^3\times\rr^3)$
for any $\phi\in\lip_b(\rr^3)$.
By Step 1, we deduce that $t \mapsto\int_{\rr^3}\int_{\rr
^3}L_{B_k}\phi(v,v_*)
f_t(dv_*)f_t(dv)$
is continuous on $[0,\infty)$.

\textit{Step 3.} We claim that
$\llvert (L_B-L_{B_k})\phi(v,v_*)\rrvert \leq C_\phi(1+\llvert v\rrvert ^2+\llvert v_*\rrvert ^2)k^{-\kappa}$
for some $\kappa=\kappa(\gamma,\nu)>0$,
for all $\phi\in\lip_b(\rr^3)$. Using (\ref{newL2}), (\ref{lit}) and
then (\ref{aaa}), we get
\begin{eqnarray*}
&&\bigl\llvert (L_B-L_{B_k})\phi(v,v_*)\bigr\rrvert \\
&&\qquad\leq
C_\phi\llvert v-v_*\rrvert ^\gamma \int_0^{\pi/2}
\int_0^{2\pi} \theta\llvert v-v_*\rrvert (
\indiq_{\{
\llvert v-v_*\rrvert ^\gamma>k\}
}+\indiq_{\{\theta\leq1/k\}})\,d\varphi b(\theta)\,d\theta
\\
&&\qquad\leq C_\phi\llvert v-v_*\rrvert ^{\gamma+1}\indiq_{\{\llvert v-v_*\rrvert ^\gamma>k\}}
+ C_\phi \llvert v-v_*\rrvert ^{\gamma+1} k^{\nu-1}
\\
&&\qquad\leq C_\phi\llvert v-v_*\rrvert ^{\gamma+1}\indiq_{\{\llvert v-v_*\rrvert ^\gamma>k\}}
+ C_\phi \bigl(1+\llvert v\rrvert ^2+\llvert v_*\rrvert
^2\bigr) k^{\nu-1}.
\end{eqnarray*}
If $\gamma\in(0,1)$, we write $\llvert v-v_*\rrvert ^{\gamma+1}\indiq_{\{
\llvert v-v_*\rrvert ^\gamma>k\}}\leq k^{1-1/\gamma}\llvert v-v_*\rrvert ^2$
and conclude with $\kappa= (1/\gamma-1)\land(1-\nu)$.
If $\gamma=0$, $\llvert v-v_*\rrvert ^\gamma>k$ never happens (since $k\geq1$),
whence the claim with
$\kappa=1-\nu$. If $\gamma\in(-1,0)$,
$\llvert v-v_*\rrvert ^\gamma>k$ implies $\llvert v-v_*\rrvert <k^{-1/\llvert \gamma\rrvert }$ and we conclude with
$\kappa= ((\gamma+1)/\llvert \gamma\rrvert )\land(1-\nu)$.

\textit{Step 4.} Let $\phi\in\lip_b(\rr^3)$. By Step 2, $L_{B_k}\phi
\in
C_b(\rr^3\times\rr^3)$
and Step 3 implies that $L_{B_k}\phi$ tends to $L_B\phi$ uniformly on
compacts, whence
$L_B\phi$ is continuous. Next, Step 3 and (\ref{energy}) show that
$\int_{\rr^3}\int_{\rr^3}L_{B_k}\phi(v,v_*) f_t(dv_*)f_t(dv)$ goes
to $\int_{\rr^3}
\int_{\rr^3}L_{B}\phi(v,v_*) f_t(dv_*)f_t(dv)$
uniformly for $t\in|0,\infty)$. Using Step 2, we conclude that
$t \mapsto\int_{\rr^3}\int_{\rr^3}L_{B}\phi(v,v_*)
f_t(dv_*)f_t(dv)$ is
continuous on $[0,\infty)$.
\end{pf}

\section{Lowerbound}\label{supp}

The aim of this section is to prove Theorem~\ref{support} and to deduce
some lowerbounds of weak solutions.
For $x\in\rr^3$ and $r>0$, we denote by $\cB(x,r):=\{y\in\rr^3 \dvtx
\llvert y-x\rrvert <r\}$
and by $\cS(x,r):=\{y\in\rr^3 \dvtx  \llvert y-x\rrvert =r\}$.
We start with the following preliminary result.

\begin{lem}\label{super}
Consider $g \in\cP(\rr^3)$ enjoying the following property:
$v_1,v_2 \in\supp g$ implies that $\cS
((v_1+v_2)/2,\llvert v_1-v_2\rrvert /2)\subset
\supp g$.
If $g$ is not a Dirac mass, then
$\supp g = \rr^3$.
\end{lem}

\begin{pf} We first claim that for any $x\in\rr^3$, any $r>0$, $\cS
(x,r)\subset\supp g$ implies
$\bar\cB(x,\sqrt2 r)\subset\supp g$. Due to our assumption, it
suffices to show that for any
$v \in\bar\cB(x,\sqrt2 r)$, there exists $v_1,v_2 \in\cS(x,r)$ such
that $v \in
\cS((v_1+v_2)/2,\llvert v_1-v_2\rrvert /2)$.
This is not hard: write $v=x+\alpha r \sigma$, for some $\sigma\in
{\mathbb{S}^2}
$ and some $\alpha\in[0,\sqrt2]$,
consider any $\tau\in{\mathbb{S}^2}$ orthogonal to $\sigma$ and choose
$v_1= x+ r[(\alpha+ \sqrt{2-\alpha^2})\sigma+(\alpha- \sqrt {2-\alpha
^2})\tau]/2$
and $v_2= x+ r[(\alpha+ \sqrt{2-\alpha^2})\sigma- (\alpha- \sqrt {2-\alpha^2})\tau]/2$.

Since $g$ is not a Dirac mass, we can find $v_1\ne v_2$ in
$\supp g$. Put $x_0=(v_1+v_2)/2$ and $r_0=\llvert v_1-v_2\rrvert /2>0$. By assumption,
$\cS(x_0,r_0)\subset\supp g$, whence $\bar\cB(x_0,\sqrt2
r_0)\subset
\supp g$. Thus in particular,
$\cS(x_0,\sqrt2 r_0)\subset\supp g$, whence\break $\bar\cB(x_0,2
r_0)\subset\supp g$, and so on.
We find that $\bar\cB(x_0,2^{n/2} r_0)\subset\supp g$ for any $n\geq1$,
which ends the proof.
\end{pf}

We can now give the proof of Theorem~\ref{support}.
Let us mention that Step 2 below is inspired by Villani \cite{v:h}, Chapter~3, Section~6.2.

\begin{pf*}{Proof of Theorem~\ref{support}}
We thus assume (\ref{aaa}) for some $\gamma\in(-1,1)$,
$\nu\in(0,1)$ and consider a weak solution $(f_t)_{t\geq0}$ to (\ref
{be}) starting from some
non-Dirac initial condition $f_0 \in\cP_2(\rr^3)$.

\textit{Step 1.} For all $t>0$, $f_t$ is not a Dirac mass. This is immediate
from the conservations of momentum and energy (\ref{energy}) and the
fact that $f_0$ is not a Dirac mass:
for all $t\geq0$,
all $v_0\in\rr^3$,
\[
\int_{\rr^3}\llvert v-v_0\rrvert ^2
f_t(dv)=\int_{\rr^3}\llvert v-v_0
\rrvert ^2 f_0(dv)>0. %
\]

\textit{Step 2.} Here we prove that for any $t>0$, any $v_0\in\rr^3$, any
$\epsilon>0$,
[recall that $v'=v'(v,v_*,\sigma)$ and $\theta=\theta(v,v_*,\sigma)$
were defined in (\ref{vprime})]
\begin{eqnarray*}
&&f_{t}\bigl(\cB(v_0,\epsilon)\bigr)=0\\
&&\quad \Longrightarrow\quad
\int_{\rr^3}\int_{\rr^3}\int
_{{\mathbb{S}^2}} \indiq_{\{
v'(v,v_*,\sigma)\in\cB
(v_0,\epsilon)\}}\\
&&\hspace*{78pt}\quad {}\times\indiq_{\{v\ne v_*,\theta(v,v_*,\sigma) \in(0,\pi/2)\}}\,d\sigma
f_{t}(dv_*)f_{t}(dv)=0. %
\end{eqnarray*}
Assume thus that $f_{t}(\cB(v_0,\epsilon))=0$ and
consider $\phi_{\epsilon,v_0}\in\lip_b(\rr^3)$, strictly positive
on $\cB
(v_0,\epsilon)$
and vanishing outside $\cB(v_0,\epsilon)$. By Lemma~\ref{trws},
$s\mapsto\int_{\rr^3}\phi_{\epsilon,v_0}(v)\*f_s(dv)$ belongs to
$C^1([0,\infty))$.
Since it is nonnegative and
vanishes at $t>0$, its derivative also vanishes at $t$. Consequently,
by (\ref{wbe}),
\[
\int_{\rr^3}\int_{\rr^3}\int
_{{\mathbb{S}^2}} B\bigl(\llvert v-v_*\rrvert ,\cos \theta\bigr) \bigl[
\phi_{\epsilon
,v_0}\bigl(v'\bigr)-\phi _{\epsilon,v_0}(v)\bigr]\,d
\sigma f_{t}(dv_*)f_{t}(dv)=0. %
\]
But $f_{t}(\cB(v_0,\epsilon))=0$ and $\supp\phi_{\epsilon
,v_0}\subset\cB(v_0,\epsilon
)$, so that
\[
\int_{\rr^3}\int_{\rr^3}\int
_{{\mathbb{S}^2}} B\bigl(\llvert v-v_*\rrvert ,\cos \theta\bigr)
\phi_{\epsilon,v_0}\bigl(v'\bigr)\,d\sigma f_{t}(dv_*)f_{t}(dv)=0.
\]
This implies the result, since $\phi_{\epsilon,v_0}(v')B(\llvert v-v_*\rrvert ,\cos
\theta
)>0$ as soon as
$v'\in\cB(v_0,\epsilon)$, $v\ne v_*$ and $\theta\in(0,\pi/2)$ due to
(\ref{aaa}).

\textit{Step 3.} We now show that for any $t>0$, $v_1,v_2\in\supp f_{t}$ implies
$\cS((v_1+v_2)/2,\llvert v_1-v_2\rrvert /2)\subset\supp f_{t}$.
We can assume that $v_1\ne v_2$, because else,
$\cS((v_1+v_2)/2,\llvert v_1-v_2\rrvert /2)=\{v_1\}$ and the result is obvious.
Observe that $\cS((v_1+v_2)/2,\llvert v_1-v_2\rrvert /2)$ is the closure of $\Delta
_{v_1,v_2} \cup\Delta_{v_2,v_1}$, where
\begin{eqnarray*}
\Delta_{v_1,v_2}:=&\bigl\{v'(v_1,v_2,
\sigma) \dvtx  \sigma\in{\mathbb{S}^2}, \theta(v_1,v_2,
\sigma) \in(0,\pi/2)\bigr\}.
\end{eqnarray*}
Since $\supp f_{t}$ is closed, it suffices to prove that
$\Delta_{v_1,v_2} \cup\Delta_{v_2,v_1} \subset\supp f_{t}$. Let thus,
for example,
$v_0 \in\Delta_{v_1,v_2}$. Then $v_0=v'(v_1,v_2,\sigma_0)$ for some
$\sigma_0 \in{\mathbb{S}^2}$
with $\theta_0=\theta(v_1,v_2,\sigma_0) \in(0,\pi/2)$.
Thus for
all $v\simeq v_1$, all $v_* \simeq v_2$, all $\sigma\simeq\sigma_0$,
we have
$v'(v,v_*,\sigma)\simeq v_0$, $v \ne v_*$ and $\theta(v,v_*,\sigma
)\in
(0,\pi/2)$.
Since $v_1 \in\supp f_{t}(dv)$ and $v_2 \in\supp f_{t}(dv_*)$, we
conclude that for any $\epsilon>0$,
\[
\int_{\rr^3}\int_{\rr^3}\int
_{{\mathbb{S}^2}} \indiq_{\{
v'(v,v_*,\sigma)\in\cB
(v_0,\epsilon)\}} \indiq_{\{v\ne v_*,\theta(v,v_*,\sigma) \in(0,\pi/2)\}}\,d\sigma
f_{t}(dv_*)f_{t}(dv)>0. %
\]
This implies that $f_{t}(\cB(v_0,\epsilon))>0$ for all $\epsilon>0$
by Step 2.

\textit{Step 4.} We conclude from Lemma~\ref{super} and Steps 1 and 3 that
for all $t>0$, $\supp f_t=\rr^3$.
\end{pf*}

We finally check the following estimate.

\begin{prop}\label{lower2} Assume (\ref{aaa}) for some
$\gamma
\in(-1,1)$,
$\nu\in(0,1)$. Let also $f_0 \in\cP_2(\rr^3)$ not be a Dirac mass.
Consider any weak solution $(f_t)_{t\geq0}$ to (\ref{be}) starting
from $f_0$.
For all $0<t_0<t_1$,
\[
q_{t_0,t_1}:=\inf_{t\in[t_0,t_1],w \in\rr^3,\zeta\in\rr^3} f_t\bigl(K(w,\zeta)
\bigr) >0, %
\]
where $K(w,\zeta):=\{v\in\rr^3 \dvtx  \llvert v\rrvert \leq3, \llvert v-w\rrvert \geq1, \llvert
\langle
v-w,\zeta \rangle\rrvert \geq\llvert \zeta\rrvert \}$.
\end{prop}

\begin{pf} We divide the proof into three steps.

\textit{Step 1.} We first prove that for any $0<t_0<t_1$,
$\inf_{t\in[t_0,t_1],x \in\cS(0,2)} f_t(\cB(x,\break 1))>0$. To this end,
consider $\phi\in\lip_b(\rr^3)$ such that $\indiq_{\cB(0,1/2)}
\leq
\phi\leq\indiq_{\cB(0,1)}$.
Define $F(t,x)=\int_{\rr^3}\phi(v-x)f_t(dv)$. We know from Lemma~\ref
{trws} that
$t \mapsto F(t,x)$ is continuous for each $x\in\rr^3$. Furthermore,
denoting by $C$ the Lipschitz
constant of $\phi$, we have $\sup_{t\geq0} \llvert F(t,x)-F(t,y)\rrvert \leq C
\llvert x-y\rrvert $. All this implies
that $F$ is continuous on $[0,\infty) \times\rr^3$. Since
$F(t,x)\geq
f_t(\cB(x,1/2))$,
we deduce from Theorem~\ref{support} that $F(t,x)>0$ for all $t>0$, all
$x\in\rr^3$.
The continuity of $F$ and the compactness of $[t_1,t_2]\times\cS(0,2)$
imply that $\inf_{[t_1,t_2]\times\cS(0,2)}F>0$. This ends the step, because
$f_t(\cB(x,1))\geq F(t,x)$.

\textit{Step 2.} Here we check that for any $w\in\rr^3$, any $\zeta\in
\rr^3$
we can find $x_{w,\zeta} \in\cS(0,2)$ such that $\cB(x_{w,\zeta
},1)\subset K(w,\zeta)$.
We may assume that $\zeta\ne0$ [because $K(w,\zeta)\subset K(w,0)$
for any $\zeta\ne0$]. Put $\sg(y)=1$ for $y \geq0$ and $\sg(y)=-1$
for $y < 0$.
Choose $x_{w,\zeta}=-2\sg( \langle w,\zeta \rangle)\zeta
/\llvert \zeta\rrvert  \in\cS(0,2)$.
It remains to prove that $\cB(x_{w,\zeta},1)\subset K(w,\zeta)$.
Let thus $v\in\cB(x_{w,\zeta},1)$.
\begin{longlist}[(b)]
\item[(a)] First, $\llvert v\rrvert \leq\llvert x_{w,\zeta}\rrvert +1= 3$.

\item[(b)] Next, observe that $\llvert w-x_{w,\zeta}\rrvert =|w+ 2\sg( \langle
w,\zeta \rangle
)\zeta
/|\zeta||\geq
\sqrt{|w|^2+4}\geq2$, so that
\[
\llvert w-v\rrvert \geq\llvert w-x_{w,\zeta}\rrvert - \llvert
x_{w,\zeta}-v\rrvert \geq2-1=1. %
\]

\item[(c)] Finally, using that $\llvert  \langle w-x_{w,\zeta},\zeta
\rangle\rrvert =
| \langle w,\zeta \rangle+ 2 \sg( \langle w,\zeta
\rangle) |\zeta|| \geq2|\zeta|$,
we see that
\[
\bigl\llvert \langle w-v,\zeta \rangle\bigr\rrvert \geq\bigl\llvert \langle
w-x_{w,\zeta},\zeta \rangle\bigr\rrvert - \bigl\llvert \langle
x_{w,\zeta
}-v,\zeta \rangle\bigr\rrvert \geq2\llvert \zeta\rrvert -\llvert
\zeta\rrvert =\llvert \zeta\rrvert . %
\]
\end{longlist}

All this shows that $v \in K(w,\zeta)$ as desired.

\textit{Step 3.} By Step 2, we have
\[
\inf_{t\in[t_0,t_1],w \in\rr^3,\zeta\in\rr^3} f_t\bigl(K(w,\zeta
)\bigr)\geq\inf_{t\in[t_0,t_1],x \in\cS(0,2)} f_t\bigl(\cB(x,1)\bigr).
\]
This last quantity is positive if $0<t_0<t_1$ by Step~1.
\end{pf}

\section{Probabilistic interpretation}\label{proba}

We write down the probabilistic interpretation of (\ref{be}) initiated
by Tanaka
\cite{t} in the case of Maxwell molecules.

\begin{prop}\label{exifc}
Let $f_0 \in\cP_2(\rr^3)$.
Assume (\ref{aaa}) for some $\gamma\in(-1,1)$, $\nu\in(0,1)$.
\begin{longlist}[(ii)]
\item[(i)] Assume first that $\gamma\in(0,1)$. Then for any weak solution
$(f_t)_{t\geq0}$ to (\ref{be}) starting from $f_0$ and satisfying
(\ref{mom}),
there exist, on some probability space $(\Omega,\cF,(\cF_t)_{t\geq
0},\Pr)$,
a $\cF_0$-measurable random variable $V_0$ with law $f_0$, a $(\cF
_t)_{t\geq0}$-Poisson measure
$N(ds,dv,d\theta,d\varphi,du)$ on $[0,\infty)\times\rr^3\times
(0,\pi
/2]\times[0,2\pi)\times[0,\infty)$
with intensity $dsf_s(dv)b(\theta)\,d\theta \,d\varphi\, du$ and a c\`adl\`ag
$(\cF_t)_{t\geq0}$-adapted $\rr^3$-valued process $(V_t)_{t\geq0}$
satisfying $\cL(V_t)=f_t$ for all $t\geq0$
and solving
%
\begin{eqnarray}
\label{bolsde}
&& V_t=V_0 + \int_0^t
\int_{\rr^3}\int_0^{\pi/2}\int
_0^{2\pi}\int_0^\infty
a(V_\sm,v,\theta,\varphi)\nonumber\\[-8pt]\\[-8pt]
&&\hspace*{143pt}{}\times\indiq_{\{u\leq\llvert V_\sm-v\rrvert ^\gamma\}
}N(ds,dv,d\theta,d
\varphi,du).\nonumber
\end{eqnarray}

\item[(ii)] Assume next that $\gamma\in(-1,0]$ and that $f_0 \in\cP_p(\rr
^3)$ for some $p>2$.
There exists a weak solution $(f_t)_{t\geq0}$ to (\ref{be}) starting
from $f_0$ satisfying
%
\begin{eqnarray}
\label{mom2} \forall T>0,\qquad \sup_{[0,T]} m_p(f_t)
\leq C_{T,p}
\end{eqnarray}
and such that there exist, on some probability space $(\Omega,\cF
,(\cF
_t)_{t\geq0},\Pr)$,
a $\cF_0$-measurable random variable $V_0$ with law $f_0$, a $(\cF
_t)_{t\geq0}$-Poisson measure
$N(ds,dv,d\theta,d\varphi,du)$ on $[0,\infty)\times\rr^3\times
(0,\pi
/2]\times[0,2\pi)\times[0,\infty)$
with intensity $dsf_s(dv)b(\theta)\,d\theta \,d\varphi \,du$ and a c\`adl\`ag
$(\cF_t)_{t\geq0}$-adapted $\rr^3$-valued process $(V_t)_{t\geq0}$
solving (\ref{bolsde})
and satisfying $\cL(V_t)=f_t$ for all $t\geq0$.
\end{longlist}
\end{prop}

The proof of this result is fastidious and not very interesting, so we
will give at the end of the paper.
In the sequel, $(V_t)_{t\geq0}$ will be called Boltzmann process.

\section{Approximation}\label{appro}

We now wish to approximate the Boltzmann process $(V_t)_{t\geq0}$ by a
process $(V_t^\epsilon)_{t\geq0}$
of which we can more easily study the law. We essentially freeze the
integrand in the Poisson
integral during a small time interval $[t-\epsilon,t]$, so that the
resulting process
$V_t^\epsilon$ becomes a L\'evy process conditionally on $\cF
_{t-\epsilon}$.
The advantage of L\'evy processes is that we can easily study their
laws through their
Fourier transforms.
Due to the lack of regularity of the function $a$, we have to make use
of $\varphi_0$ introduced
in Lemma~\ref{tanana}.

\begin{prop}\label{app}
Assume (\ref{aaa}) for some $\gamma\in(-1,1)$, $\nu\in(0,1)$
with $\gamma+\nu>0$.
Consider a Boltzmann process $(V_t)_{t\geq0}$
built with a Poisson measure $N$ as in Proposition~\ref{exifc}. For
$\epsilon
\in(0,t\land1)$, set
%
\begin{eqnarray}
\label{vte}
&&
V_t^\epsilon:=V_{t-\epsilon} + \int
_{t-\epsilon}^t \int_{\rr
^3}\int
_0^{\pi/2}\int_0^{2\pi}
\int_0^\infty a\bigl(V_{t-\epsilon},v,\theta,
\varphi+\varphi_0(V_\sm-v,V_{t-\epsilon
}-v)\bigr)
\nonumber\\[-8pt]\\[-8pt]
&&\hspace*{166pt}{}\times\indiq_{\{u\leq\llvert V_{t-\epsilon}-v\rrvert ^\gamma\}}N(ds,dv,d\theta ,d\varphi ,du).
\nonumber
\end{eqnarray}
\begin{longlist}[(ii)]
\item[(i)] If $\gamma\in(0,1)$, then for any $0<t_0 \leq t-\epsilon\leq t$ with
$\epsilon\in(0,1)$ and
any $\eta\in(0,2)$,
\[
\E \bigl[\bigl\llvert V_t-V_t^\epsilon\bigr
\rrvert ^\nu \bigr] \leq C_{t_0,\eta} \epsilon ^{2-\eta}.
\]

\item[(ii)] If $\gamma\in(-1,0]$, then for any $0 \leq t-\epsilon\leq t $
with $\epsilon
\in(0, 1)$ and
any $\eta\in(0,2+\gamma/\nu)$,
\[
\E \bigl[\bigl\llvert V_t-V_t^\epsilon\bigr
\rrvert ^\nu \bigr] \leq C_{\eta} \epsilon ^{2+\gamma/\nu-\eta}.
\]
\end{longlist}
\end{prop}

We will use that for $a,b>0$, there are some constants $0<c_{a,b}<C_{a,b}$
such that
%
\begin{eqnarray}
\label{thf}
\forall x,y>0,\qquad c_{a,b}\bigl\llvert x^{a+b}-y^{a+b}
\bigr\rrvert &\leq&\bigl(x^a+y^a\bigr)\bigl\llvert
x^b-y^b\bigr\rrvert \nonumber\\[-8pt]\\[-8pt]&\leq &C_{a,b}\bigl\llvert
x^{a+b}-y^{a+b}\bigr\rrvert .\nonumber
\end{eqnarray}

\begin{pf} We divide the proof into several steps.

\textit{Step 1.} Here we check that for all $\beta\in(\nu,1)$ and all $0
\leq s \leq t$,
$\E[\llvert V_t-V_s\rrvert ^\beta]\leq C_\beta(t-s)$ in both cases (i) and (ii).
Using the subadditivity of $x \mapsto x^\beta$, we deduce from (\ref
{bolsde}) that
\begin{eqnarray*}
&&\llvert V_t-V_s\rrvert ^\beta\leq\int
_s^t \int_{\rr^3}\int
_0^{\pi/2} \int_0^{2\pi
}
\int_0^\infty \bigl\llvert a(V_{r-},v,
\theta,\varphi)\bigr\rrvert ^\beta \\
&&\hspace*{155pt}{}\times\indiq_{\{u\leq\llvert V_{r-}-v\rrvert ^\gamma\}} N(dr,dv,d
\theta,d\varphi,du). %
\end{eqnarray*}
Taking expectations, integrating in $u$ and using (\ref{lit}), we obtain
\begin{eqnarray*}
\E\bigl[\llvert V_t-V_s\rrvert ^\beta\bigr]
&\leq& \E \biggl[\int_s^t \int
_{\rr^3}\int_0^{\pi/2} \int
_0^{2\pi}\int_0^\infty
\bigl\llvert a(V_{r},v,\theta,\varphi)\bigr\rrvert ^\beta
\\
&&\hspace*{107pt}{}\times\indiq_{\{u\leq\llvert V_{r}-v\rrvert ^\gamma\}}\, du \,d\varphi b(\theta)\,d\theta f_r(dv)\,dr
\biggr]
\\
&\leq& \int_s^t \int_{\rr^3}
\int_0^{\pi/2} \int_0^{2\pi}
\theta ^\beta \E \bigl[\llvert V_r-v \rrvert
^{\gamma+\beta}\bigr]\,d\varphi b(\theta)\,d\theta f_r(dv)\,dr
\\
&\leq& C_\beta(t-s).
\end{eqnarray*}
We used that $\beta>\nu$, whence $\int_0^{\pi/2} \theta^\beta
b(\theta
)\,d\theta\leq
C_0 \int_0^{\pi/2} \theta^{\beta-1-\nu}\,d\theta<\infty$ by
(\ref{aaa}),
that $\llvert V_r-v \rrvert ^{\gamma+\beta}\leq C(1+\llvert V_r\rrvert ^2+\llvert v\rrvert ^2)$ [because
$\gamma
+\beta\in(0,2)$] and that
$\int_{\rr^3}\E(1+\llvert v\rrvert ^2+\llvert V_r\rrvert ^2) f_r(dv)=1+2m_2(f_r)=C$ by (\ref{energy})
[recall that $\cL(V_t)=f_t$].

\textit{Step 2.} In this step we prove that for all $\beta\in(\nu,1)$
and all $0 \leq t-\epsilon\leq t$, in
cases (i) and (ii),
\begin{eqnarray*}
\E\bigl[\bigl\llvert V_t-V_t^\epsilon\bigr\rrvert
^\beta\bigr]\leq C_\beta\int_{t-\epsilon}^t
\int_{\rr^3}\E\bigl[A^{1,\beta
,\epsilon}_s(v) +
A^{2,\beta,\epsilon}_s(v) +A^{3,\beta,\epsilon}_s(v)\bigr]
f_s(dv)\,ds,
\end{eqnarray*}
where, using the notation $x_+=x \lor0$,
\begin{eqnarray*}
A^{1,\beta,\epsilon}_s(v)& :=& \bigl(\llvert V_{t-\epsilon}-v\rrvert
^\gamma\land \llvert V_{s}-v\rrvert ^\gamma \bigr)\\
&&{}\times
\bigl(\llvert V_s-V_{t-\epsilon}\rrvert ^\beta\land
\bigl[\llvert V_{t-\epsilon}-v\rrvert ^\beta+ \llvert
V_{s}-v\rrvert ^\beta\bigr] \bigr),
\\
A^{2,\beta,\epsilon}_s(v) &:=& \bigl(\llvert V_{t-\epsilon}-v\rrvert
^\gamma- \llvert V_{s}-v\rrvert ^\gamma \bigr)_+
\llvert V_{t-\epsilon}-v\rrvert ^\beta,
\\
A^{3,\beta,\epsilon}_s(v)& :=& \bigl(\llvert V_{s}-v\rrvert
^\gamma- \llvert V_{t-\epsilon}-v\rrvert ^\gamma \bigr)_+
\llvert V_s-v\rrvert ^\beta.
\end{eqnarray*}
Exactly as in Step 1, we obtain
\begin{eqnarray}
&&\hspace*{-20pt}\E\bigl[\bigl\llvert V_t-V_t^\epsilon\bigr\rrvert
^\beta\bigr]\nonumber\\
&&\hspace*{-20pt}\qquad\leq \E \biggl[\int_{t-\epsilon}^t
\int_{\rr^3}\int_0^{\pi/2} \int
_0^{2\pi}\int_0^\infty
\bigl\llvert a(V_s,v,\theta,\varphi)\indiq_{\{u\leq\llvert V_s-v\rrvert ^\gamma\}}\nonumber
\\
&&\hspace*{133pt}{} -a\bigl(V_{t-\epsilon},v,\theta,\varphi+\varphi_0(V_s-v,V_{t-\epsilon
}-v)\bigr)\nonumber\\
\eqntext{{}\times\indiq_{\{
u\leq\llvert V_{t-\epsilon}-v\rrvert ^\gamma\}} \bigr\rrvert ^\beta \,du \,d\varphi b(\theta)
\,d\theta f_s(dv)\,ds \biggr].}
\end{eqnarray}
Integrating in $u$, we get
$\E[\llvert V_t-V_t^\epsilon\rrvert ^\beta]\leq\int_{t-\epsilon}^t \int_{\rr^3}
\E[B^{1,\beta,\epsilon}_s(v) + B^{2,\beta,\epsilon
}_s(v)+B^{2,\beta,\epsilon}_s(v)]
f_s(dv)\,ds$,
where
\begin{eqnarray*}
B^{1,\beta,\epsilon}_s(v)&:=&\int_0^{\pi/2}
\int_0^{2\pi} \bigl(\llvert V_{t-\epsilon}-v
\rrvert ^\gamma\land\llvert V_s-v\rrvert ^\gamma
\bigr)
\\
&&\hspace*{41pt}{}\times \bigl\llvert a(V_s,v,\theta,\varphi)\\
&&\hspace*{54pt}{}-a\bigl(V_{t-\epsilon},v,
\theta,\varphi +\varphi _0(V_s-v,V_{t-\epsilon}-v)
\bigr)\bigr\rrvert ^\beta\, d\varphi b(\theta)\,d\theta,
\\
B^{2,\beta,\epsilon}_s(v)&:=& \int_0^{\pi/2}
\int_0^{2\pi} \bigl(\llvert V_{t-\epsilon
}-v
\rrvert ^\gamma- \llvert V_s-v\rrvert ^\gamma
\bigr)_+\\
&&\hspace*{45pt}{}\times \bigl\llvert a\bigl(V_{t-\epsilon},v,\theta,\varphi+
\varphi_0(V_s-v,V_{t-\epsilon
}-v)\bigr)\bigr\rrvert
^\beta \,d\varphi b(\theta)\,d\theta,
\\
B^{3,\beta,\epsilon}_s(v)&:=&\int_0^{\pi/2}
\int_0^{2\pi} \bigl(\llvert V_s-v
\rrvert ^\gamma - \llvert V_{t-\epsilon}-v\rrvert ^\gamma
\bigr)_+ \bigl\llvert a(V_s,v,\theta,\varphi)\bigr\rrvert
^\beta \,d\varphi b(\theta)\,d\theta.
\end{eqnarray*}
Using Lemma~\ref{tanana} and (\ref{lit}), we realize that
\begin{eqnarray*}
&&\bigl\llvert a(V_s,v,\theta,\varphi)-a\bigl(V_{t-\epsilon},v,
\theta,\varphi +\varphi _0(V_s-v,V_{t-\epsilon}-v)
\bigr)\bigr\rrvert\\
&&\qquad \leq2 \theta \bigl(\llvert V_s-V_{t-\epsilon}
\rrvert \land\bigl[\llvert V_{t-\epsilon}-v\rrvert + \llvert
V_{s}-v\rrvert \bigr] \bigr). %
\end{eqnarray*}
Since $\int_0^{\pi/2} \theta^\beta b(\theta)\,d\theta<\infty$,
we deduce that $B^{1,\beta,\epsilon}_s(v) \leq C_\beta A^{1,\beta
,\epsilon}_s(v)$.
Using (\ref{lit}),
we get $B^{2,\beta,\epsilon}_s(v) \leq C_\beta A^{2,\beta,\epsilon
}_s(v)$ and
$B^{3,\beta,\epsilon}_s(v) \leq
C_\beta A^{3,\beta,\epsilon}_s(v)$,
which completes the step.

\textit{Step 3.} Here we conclude the proof of (i). We thus assume that
$\gamma\in(0,1)$ and fix
$0<t_0\leq t-\epsilon\leq t$ with $\epsilon\in(0,1)$. We also fix
$\beta\in(\nu
,1)$ and apply Step 2. We first
observe that
\[
A^{1,\beta,\epsilon}_s(v) \leq C\bigl(\llvert v\rrvert ^\gamma+
\llvert V_{t-\epsilon}\rrvert ^\gamma +\llvert V_s\rrvert
^\gamma \bigr) \llvert V_s-V_{t-\epsilon}\rrvert
^\beta. %
\]
We next use twice (\ref{thf}) (with $a=\gamma$ and $b=\beta$) to
deduce that
\begin{eqnarray*}
A^{2,\beta,\epsilon}_s(v)+A^{3,\beta,\epsilon}_s(v) &\leq& \bigl(
\llvert V_{t-\epsilon}-v\rrvert ^\beta+\llvert V_s-v
\rrvert ^\beta\bigr) \bigl\llvert \llvert V_{t-\epsilon
}-v\rrvert
^\gamma- \llvert V_s-v\rrvert ^\gamma\bigr\rrvert
\\
&\leq&C_\beta\bigl\llvert \llvert V_{t-\epsilon}-v\rrvert
^{\beta+\gamma} - \llvert V_s-v\rrvert ^{\beta+\gamma
}\bigr\rrvert
\\
&\leq& C_\beta\bigl(\llvert V_{t-\epsilon}-v\rrvert
^\gamma+\llvert V_s-v\rrvert ^\gamma\bigr) \bigl
\llvert \llvert V_{t-\epsilon
}-v\rrvert ^\beta- \llvert
V_s-v\rrvert ^\beta\bigr\rrvert
\\
&\leq& C_\beta\bigl(\llvert V_{t-\epsilon}-v\rrvert
^\gamma+\llvert V_s-v\rrvert ^\gamma \bigr)\llvert
V_s-V_{t-\epsilon}\rrvert ^\beta
\\
&\leq& C_\beta\bigl(\llvert v\rrvert ^\gamma+\llvert
V_{t-\epsilon}\rrvert ^\gamma+\llvert V_s\rrvert
^\gamma\bigr) \llvert V_s-V_{t-\epsilon}\rrvert
^\beta.
\end{eqnarray*}
We thus have
\begin{eqnarray*}
\E\bigl[\bigl\llvert V_t-V_t^\epsilon\bigr\rrvert
^\beta\bigr] &\leq& C_\beta\int_{t-\epsilon}^t
\int_{\rr^3}\E \bigl[ \llvert V_s-V_{t-\epsilon}
\rrvert ^\beta \bigl(\llvert v\rrvert ^\gamma+\llvert
V_{t-\epsilon}\rrvert ^\gamma+ \llvert V_s\rrvert
^\gamma\bigr) \bigr]f_s(dv)\,ds
\\
&\leq& C_\beta\int_{t-\epsilon}^t \E \bigl[
\llvert V_s-V_{t-\epsilon
}\rrvert ^\beta\bigl(1+\llvert
V_{t-\epsilon
}\rrvert ^\gamma+ \llvert V_s\rrvert
^\gamma\bigr) \bigr]\,ds,
\end{eqnarray*}
since $\int_{\rr^3}\llvert v\rrvert ^\gamma f_s(dv) \leq\int_{\rr^3}(1+\llvert v\rrvert ^2)
f_t(dv) \leq C$
by (\ref{energy}).
We now consider $\delta\in(0,1-\beta)$ and apply the H\"older inequality
[with $p=1/(1-\delta)$ and $q=1/\delta$]:
\begin{eqnarray*}
&&\E\bigl[\bigl\llvert V_t-V_t^\epsilon\bigr\rrvert
^\beta\bigr] \leq C_\beta\int_{t-\epsilon}^t
\E \bigl[ \llvert V_s-V_{t-\epsilon
}\rrvert ^{\beta/(1-\delta
)}
\bigr]^{1-\delta} \\[-2pt]
&&\hspace*{110pt}{}\times\E \bigl[ \bigl(1+\llvert V_{t-\epsilon}\rrvert
^\gamma+\llvert V_s\rrvert ^\gamma
\bigr)^{1/\delta} \bigr]^{\delta}\,ds. %
\end{eqnarray*}
By Step 1 [observe that $\beta/(1-\delta) \in(\nu,1)$], we have $\E[
\llvert V_s-V_{t-\epsilon}\rrvert ^{\beta/(1-\delta)}]\leq
C_{\beta,\delta}\epsilon$
for all $s\in[t-\epsilon,t]$.
Using (\ref{mom}) [recall that
$\cL(V_s)=f_s$ for all $s\geq0$], we see that
$\E[ (1+\llvert V_{t-\epsilon}\rrvert ^\gamma+\llvert V_s\rrvert ^\gamma)^{1/\delta}] \leq
C_{t_0,\delta
}$ (because $s\geq t-\epsilon\geq t_0>0$). Thus
\[
\E\bigl[\bigl\llvert V_t-V_t^\epsilon\bigr\rrvert
^\beta\bigr] \leq C_{\beta,\delta,t_0} \int_{t-\epsilon}^t
\epsilon ^{1-\delta}\,ds \leq C_{\beta,\delta,t_0} \epsilon^{2-\delta}.
\]
Using finally the H\"older inequality, we deduce that for all $\beta
\in(\nu,1)$ and all
$\delta\in(0,1-\beta)$,
$\E[\llvert V_t-V_t^\epsilon\rrvert ^\nu] \leq\E[\llvert V_t-V_t^\epsilon\rrvert ^\beta
]^{\nu/\beta} \leq
C_{\beta,\delta,t_0} \epsilon^{(2-\delta)\nu/\beta}$.
Since we can choose $\beta\in(\nu,1)$ arbitrarily close to $\nu$
and $\delta\in(0,1-\beta)$ arbitrarily close to $0$, it holds that
$(2-\delta)\nu/\beta\in(0,2)$
is arbitrarily close to $2$, which ends the proof of (i).

\textit{Step 4.} We finally check (ii). We thus assume that $\gamma\in
(-1,0]$, that $\gamma+\nu>0$
and we fix $0\leq t-\epsilon\leq t$ with $\epsilon\in(0,1)$.
We also fix $\beta\in(\nu,1)$ and apply Step 2. First, since
$\llvert \gamma
\rrvert /\beta\in(0,1)$,
\begin{eqnarray*}
A^{1,\beta,\epsilon}_s(v) &\leq& \bigl(\llvert V_{t-\epsilon}-v\rrvert
^\gamma\land \llvert V_{s}-v\rrvert ^\gamma\bigr)
\llvert V_s-V_{t-\epsilon}\rrvert ^{\beta(1-\llvert \gamma\rrvert /\beta)} \\[-2pt]
&&{}\times\bigl(\llvert
V_{t-\epsilon}-v\rrvert ^\beta+\llvert V_{s}-v\rrvert
^\beta\bigr)^{\llvert \gamma\rrvert /\beta}
\\[-2pt]
&\leq& \bigl(\llvert V_{t-\epsilon}-v\rrvert ^\gamma\land\llvert
V_{s}-v\rrvert ^\gamma \bigr) \bigl(\llvert
V_{t-\epsilon
}-v\rrvert ^{\llvert \gamma\rrvert }+\llvert V_{s}-v\rrvert
^{\llvert \gamma\rrvert }\bigr) \llvert V_s-V_{t-\epsilon}\rrvert
^{\beta+\gamma}
\\[-2pt]
&\leq& 2\llvert V_s-V_{t-\epsilon}\rrvert ^{\beta+\gamma}.
\end{eqnarray*}
Next, using twice (\ref{thf}) with $a=\llvert \gamma\rrvert $ and $b=\beta+\gamma$
(lines 2 and 4),
\begin{eqnarray*}
A^{2,\beta,\epsilon}_s(v)& =& \indiq_{\{\llvert V_{t-\epsilon}-v\rrvert <
\llvert V_{s}-v\rrvert \}} \bigl(\llvert
V_{s}-v\rrvert ^{\llvert \gamma\rrvert } -\llvert V_{t-\epsilon}-v\rrvert
^{\llvert \gamma\rrvert }\bigr) \llvert V_{t-\epsilon}-v\rrvert ^{\beta+\gamma} \llvert
V_{s}-v\rrvert ^{\gamma}
\\[-2pt]
&\leq& C_\beta\indiq_{\{\llvert V_{t-\epsilon}-v\rrvert < \llvert V_{s}-v\rrvert \}} \bigl(\llvert V_{s}-v
\rrvert ^{\beta} -\llvert V_{t-\epsilon}-v\rrvert ^{\beta}\bigr)
\llvert V_{s}-v\rrvert ^{\gamma}
\\[-2pt]
&\leq& C_\beta\indiq_{\{\llvert V_{t-\epsilon}-v\rrvert < \llvert V_{s}-v\rrvert \}} \frac
{\llvert V_{s}-v\rrvert ^{\beta} -\llvert V_{t-\epsilon}-v\rrvert ^{\beta}} {
\llvert V_{s}-v\rrvert ^{\llvert \gamma\rrvert } + \llvert V_{t-\epsilon}-v\rrvert ^{\llvert \gamma\rrvert }}
\\[-2pt]
&\leq& C_\beta\bigl(\llvert V_s-v\rrvert ^{\beta+\gamma}-
\llvert V_{t-\epsilon}-v\rrvert ^{\beta
+\gamma}\bigr)
\\[-2pt]
&\leq& C_\beta\llvert V_s-V_{t-\epsilon}\rrvert
^{\beta+\gamma},
\end{eqnarray*}
where we finally used that $0<\beta+\gamma<1$.
Treating $A^{3,\beta,\epsilon}_s(v)$ similarly, we finally get
\begin{eqnarray*}
\E\bigl[\bigl\llvert V_t-V_t^\epsilon\bigr\rrvert
^\beta\bigr]&\leq& C_\beta\int_{t-\epsilon}^t
\int_{\rr^3}\E \bigl[ \llvert V_s-V_{t-\epsilon
}
\rrvert ^{\beta+\gamma} \bigr] f_s(dv)\,ds\\[-2pt]a
& \leq& C_\beta\int
_{t-\epsilon}^t \E \bigl[ \llvert V_s-V_{t-\epsilon
}
\rrvert ^{\beta+\gamma} \bigr]\,ds.
\end{eqnarray*}
Using the H\"older inequality (recall that $0<\beta+\gamma<\beta$) and
Step 1, we obtain
\begin{eqnarray*}
\E\bigl[\bigl\llvert V_t-V_t^\epsilon\bigr\rrvert
^\beta\bigr] \leq C_\beta\int_{t-\epsilon}^t
\E \bigl[ \llvert V_s-V_{t-\epsilon
}\rrvert ^{\beta}
\bigr]^{1+\gamma/\beta}\,ds \leq C_\beta\epsilon^{2+\gamma/\beta},
\end{eqnarray*}
whence $\E[\llvert V_t-V_t^\epsilon\rrvert ^\nu]\leq\E[\llvert V_t-V_t^\epsilon\rrvert ^\beta
]^{\nu/\beta}
\leq C_\beta\epsilon^{(2+\gamma/\beta)\nu/\beta}$.
Since we can choose $\beta\in(\nu,1)$ arbitrarily close to $\nu$
it holds that $(2+\gamma/\beta)\nu/\beta\in(0,2+\gamma/\nu)$ is
arbitrarily close to
$2+\gamma/\nu$, which completes the proof of (ii).
\end{pf}

\section{Density estimate for the approximate process}\label{reglev}

The aim of this section, strongly inspired by Schilling, Sztonyk and Wang
\cite{ssw}, Propositions 2.1, 2.2, 2.3,
is to prove that $V^\epsilon_t$ has a regular law in some sense, with some
precise estimates in terms of $\epsilon$.

\begin{prop}\label{estibol3} Assume (\ref{aaa}) for some
$\gamma
\in(-1,1)$,
$\nu\in(0,1)$. Let $f_0 \in\cP_2(\rr^3)$ not be a Dirac mass.
If $\gamma\in(-1,0]$, assume additionally that $f_0 \in\cP
_{4+\gamma
+4\llvert \gamma\rrvert /\nu}(\rr^3)$.
Consider the approximate Boltzmann process $V^\epsilon_t$ defined in
Proposition~\ref{app} associated
with a weak solution $(f_t)_{t\geq0}$ to (\ref{be}) starting from $f_0$.
For all $h \in\rr^d$, all $\phi\in L^\infty(\rr^3)$, all
$0<t_0\leq t-\epsilon<t \leq t_1$ with $\epsilon\in(0,1)$,
\[
\bigl\llvert \E \bigl[\phi\bigl(V_t^\epsilon+h\bigr)-\phi
\bigl(V^\epsilon_t\bigr) \bigr]\bigr\rrvert \leq
C_{t_0,t_1} \llVert \phi\rrVert _{L^\infty(\rr^3)} \frac{\llvert h\rrvert }{\epsilon^{1/\nu}}.
\]
\end{prop}

We will use the following easy estimate, which resembles \cite{ssw}, Proposition~2.1:
it is much less general, but sharper.

\begin{lem}\label{ssw1}
Let $\lambda$ be a nonnegative measure on $\rr^3$ such that $\int_{\rr^3}
\llvert y\rrvert\* \lambda(dy)<\infty$
and consider
the infinitely divisible distribution $k$ with Fourier transform
\begin{eqnarray*}
\hat k(\xi):=\int_{\rr^3}e^{i \langle\xi,x \rangle
}k(dx)=\exp\bigl(-
\Phi(\xi)\bigr) \qquad\mbox{with } \Phi(\xi)=\int_{\rr^3}
\bigl(1-e^{i \langle\xi
,y \rangle} \bigr) \lambda(dy).
\end{eqnarray*}
If the right-hand side of the following inequality is finite, then $k$ has a
density (still denoted
by $k$) and
\[
\llVert \nabla k\rrVert _{L^1(\rr^3)} \leq C \bigl(1+m_1^4(
\lambda)+m_4(\lambda) \bigr) \int_{\rr^3}
e^{- \Re\Phi(\xi)} \bigl(1+\llvert \xi\rrvert \bigr)\,d\xi, %
\]
where $m_n(\lambda)=\int_{\rr^3}\llvert y\rrvert ^n \lambda(dy)$ and $C$ is a
universal constant.
\end{lem}

\begin{pf} The proof is quite similar to \cite{ssw}, Proposition~2.1.
We will show that
%
\begin{eqnarray}\label{td1}
\hspace*{15pt}\llVert \nabla k \rrVert _{L^\infty(\rr^3)} &\leq& C \int_{\rr^3}
e^{- \Re\Phi(\xi)} \llvert \xi\rrvert \,d\xi,
\\
\label{td2}\hspace*{15pt}
\bigl\llVert \llvert x\rrvert ^4\nabla k(x)\bigr\rrVert
_{L^\infty(\rr^3)} &\leq& C \bigl(1+m_1^4(\lambda
)+m_4(\lambda) \bigr) \int_{\rr^3}e^{- \Re\Phi(\xi)}
\bigl(1+\llvert \xi\rrvert \bigr)\,d\xi,
\end{eqnarray}
from which the result follows, since $(1+\llvert x\rrvert )^{-4}\in L^1(\rr^3)$. First,
\begin{eqnarray*}
\llVert \nabla k \rrVert _{L^\infty(\rr^3)} &\leq&(2\pi)^{-3}\llVert
\widehat{\nabla k}\rrVert _{L^1(\rr^3)} =(2\pi)^{-3}\bigl\llVert \xi
\hat k(\xi)\bigr\rrVert _{L^1(\rr^3)}\\
&=& (2\pi)^{-3}\int
_{\rr^3}e^{-
\Re\Phi(\xi)} \llvert \xi\rrvert \,d\xi, %
\end{eqnarray*}
whence (\ref{td1}).
To check (\ref{td2}), we start with
\[
\bigl\llVert \llvert x\rrvert ^4\nabla k(x)\bigr\rrVert
_{L^\infty(\rr^3)}\leq(2\pi)^{-3}\bigl\llVert \Delta ^2(
\widehat{\nabla k})\bigr\rrVert _{L^1(\rr^3)} \leq C \bigl\llVert
D^4\bigl(\xi\hat k(\xi)\bigr)\bigr\rrVert _{L^1(\rr^3)}.
\]
A tedious computation recalling that $\hat k(\xi)=e^{-\Phi(\xi)}$
shows that
\begin{eqnarray*}
&&\bigl\llvert D^4 \bigl(\xi\hat k(\xi)\bigr)\bigr\rrvert \\
&&\qquad\leq C
\bigl(1+\llvert \xi\rrvert \bigr)\bigl\llvert e^{-\Phi(\xi)}\bigr\rrvert \\
&&\hspace*{29pt}{}\times\bigl(
\bigl\llvert D^4 \Phi (\xi)\bigr\rrvert +\bigl\llvert D^3
\Phi(\xi) D \Phi(\xi)\bigr\rrvert + \bigl\llvert D^2\Phi(\xi)\bigr
\rrvert ^2+ \bigl\llvert D\Phi(\xi)\bigr\rrvert ^2 \bigl
\llvert D^2\Phi(\xi)\bigr\rrvert
\\
&&\hspace*{76pt}{} +\bigl\llvert D \Phi(\xi)\bigr\rrvert ^4+\bigl\llvert
D^3 \Phi(\xi)\bigr\rrvert +\bigl\llvert D\Phi(\xi)\bigr\rrvert \bigl
\llvert D^2\Phi(\xi)\bigr\rrvert +\bigl\llvert D\Phi (\xi)\bigr\rrvert
^3 \bigr).
\end{eqnarray*}
But from the expression of $\Phi$, we see that $\llvert D^n \Phi(\xi)\rrvert  \leq
m_n(\lambda)$ for all $n\geq1$.
Since $\llvert e^{-\Phi(\xi)}\rrvert =e^{-\Re\Phi(\xi)}$, we get, setting
$m_n=m_n(\lambda)$ for simplicity,
\begin{eqnarray*}
\bigl\llvert D^4 \bigl(\xi\hat k(\xi)\bigr)\bigr\rrvert &\leq& C
\bigl(1+\llvert \xi\rrvert \bigr)e^{-\Re\Phi(\xi)} \\
&&{}\times\bigl(m_4+m_3m_1+m_2^2+m_1^2m_2+
m_1^4+m_3+m_1m_2+m_1^3
\bigr)
\\
&\leq& C\bigl(1+\llvert \xi\rrvert \bigr)e^{-\Re\Phi(\xi)} \bigl(1+m_4+m_3^{4/3}+m_2^2+m_1^4
\bigr),
\end{eqnarray*}
where we used the Young inequality.
To complete the proof of (\ref{td2}), it only remains to check that
$m_3^{4/3}+m_2^2\leq
C(m_4+m_1^4)$, which is not hard by the H\"older and Young inequalities.
\end{pf}

Unfortunately, applying directly Lemma~\ref{ssw1}
to the law of $V^\epsilon_t$ does not give the correct power of
$\epsilon$.
We thus use the same trick as in \cite{ssw}: we only consider the part
of $V^\epsilon_t$ corresponding to
small values of $\theta$ (grazing collisions), in such a way that
it does not affect the estimate from below of $\Re\Phi(\xi)$,
but which makes consequently decrease the moment estimates [of
$m_1^4(\lambda)+m_4(\lambda)$].

We start with the following remark.

\begin{lem}\label{psibol} Adopt the notation and assumptions of
Proposition~\ref{estibol3}.
Let $\epsilon\in(0,t\land1)$ be fixed.
\begin{longlist}[(iii)]
\item[(i)] We can find a $(\cF_t)_{t\geq0}$-Poisson measure $M$ with the same
intensity as $N$ (see Proposition~\ref{exifc}) such that
%
\begin{eqnarray}
\label{vte2}
 V_t^\epsilon&:=&V_{t-\epsilon} + \int
_{t-\epsilon}^t \int_{\rr
^3}\int
_0^{\pi/2}\int_0^{2\pi}
\int_0^\infty a(V_{t-\epsilon},v,\theta,
\varphi)\nonumber\\[-8pt]\\[-8pt]
&&\hspace*{137pt}{}\times\indiq_{\{u\leq\llvert V_{t-\epsilon
}-v\rrvert ^\gamma\}
}M(ds,dv,d\theta,d\varphi,du).\nonumber
\end{eqnarray}

\item[(ii)] We write $V^\epsilon_t=U^\epsilon_t+W^\epsilon_t$ with
\begin{eqnarray*}
U_t^\epsilon&:=& \int_{t-\epsilon}^t
\int_{\rr^3}\int_0^{\pi
/2}\int
_0^{2\pi}\int_0^\infty
a(V_{t-\epsilon},v,\theta,\varphi)\\
&&\hspace*{100pt}{}\times\indiq_{\{u\leq\llvert V_{t-\epsilon
}-v\rrvert ^\gamma\}} \indiq_{\{\theta<\epsilon^{1/\nu}\}}
M(ds,dv,d\theta,d\varphi,du),
\\
W_t^\epsilon&:=&V_{t-\epsilon} + \int_{t-\epsilon}^t
\int_{\rr
^3}\int_0^{\pi/2}\int
_0^{2\pi}\int_0^\infty
a(V_{t-\epsilon},v,\theta,\varphi)\indiq_{\{u\leq\llvert V_{t-\epsilon
}-v\rrvert ^\gamma\}}\\
&&\hspace*{136pt}{}\times \indiq_{\{\theta\geq\epsilon^{1/\nu}\}}
M(ds,dv,d\theta,d\varphi,du),
\end{eqnarray*}
so that $U^\epsilon_t$ and $W^\epsilon_t$ are independent
conditionally on $\cF
_{t-\epsilon}$.

\item[(iii)] For all $\xi\in\rr^3$,
$\E[ e^{i  \langle\xi, U^\epsilon_t  \rangle
} \rrvert  \cF
_{t-\epsilon}  ]=
\exp (- \Psi_{\epsilon,t,V_{t-\epsilon}}(\xi)  )$,
where, for $v_0\in\rr^3$,
\[
\Psi_{\epsilon,t,v_0}(\xi)=\int_{t-\epsilon}^t \int
_{\rr^3}\int_0^{\epsilon^{1/\nu}} \int
_0^{2\pi} \bigl(1 - e^{i  \langle\xi, a(v_0,v,\theta
,\varphi)  \rangle} \bigr)
\llvert v-v_0\rrvert ^\gamma \,d\varphi b(\theta)\,d\theta
f_s(dv)\,ds. %
\]
\end{longlist}
\end{lem}

\begin{pf}
To prove point (i), define $M$ as the image measure of $N$ by the $(\cF
_t)_{t\geq0}$-predictable map
$(s,v,\theta,\varphi,u) \mapsto(s,v,\theta,\varphi+\varphi
_0(V_\sm
-v,V_{t-\epsilon}-v)$ modulo $2\pi,u)$.
Then (\ref{vte}) obviously rewrites as (\ref{vte2}). The fact that
$M$ is a $(\cF_t)_{t\geq0}$-Poisson measure with the same intensity as
$N$ is due to the fact that
the Lebesgue measure on $[0,2\pi)$ is invariant by translation (modulo
$2\pi$). This was already
noticed by Tanaka \cite{t}; see \cite{fsm}, Lemma~4.7, for a very
similar statement.
Points (ii) and (iii) follow from standard properties of Poisson
measures, because
in $U^\epsilon_t$ and $W^\epsilon_t$, the integrands are $\cF
_{t-\epsilon}$-measurable
and the Poisson integrals concern
the time interval $[t-\epsilon,t]$.
\end{pf}

We next estimate the Fourier transform of the law of $U^\epsilon_t$.

\begin{lem}\label{estibol} Adopt the notation and assumptions of
Proposition~\ref{estibol3}.
Recall that
$\Psi_{\epsilon,t,v_0}$ was defined in Lemma~\ref{psibol}. For all
$\xi\in
\rr^3$, all
$0<t_0\leq t-\epsilon<t \leq t_1$ with $\epsilon\in(0,1)$,
\[
\Re\Psi_{\epsilon,t,v_0} \bigl(\epsilon^{-1/\nu}\xi\bigr) \geq \cases{
\displaystyle c_{t_0,t_1} \bigl(\llvert \xi\rrvert
^2 \land\llvert \xi\rrvert ^\nu\bigr) & \quad if
$\gamma\in (0,1)$,
\cr\displaystyle
c_{t_0,t_1} \bigl(1+\llvert v_0\rrvert \bigr)^\gamma
\bigl(\llvert \xi\rrvert ^2 \land\llvert \xi\rrvert ^\nu
\bigr)&\quad if  $\gamma\in(-1,0]$.} %
\] %
\end{lem}

\begin{pf} We divide the proof into three steps.

\textit{Step 1.} Here we assume that $\gamma\in(-1,1)$. We have %
\begin{eqnarray*}
\Re
\Psi_{\epsilon,t,v_0} \bigl(\epsilon^{-1/\nu}\xi\bigr)&=&\int
_{t-\epsilon
}^t \int_{\rr^3}\int
_0^{\epsilon
^{1/\nu}} \int_0^{2\pi}
\bigl(1 - \cos\bigl(\epsilon^{-1/\nu} \bigl\langle\xi, a(v_0,v,
\theta ,\varphi) \bigr\rangle\bigr) \bigr)\\
&&\hspace*{82pt}{}\times \llvert v-v_0\rrvert
^\gamma \,d\varphi b(\theta)\,d\theta f_s(dv)\,ds. %
\end{eqnarray*}
By (\ref{dfvprime}),
$ \langle\xi, a(v_0,v,\theta,\varphi)  \rangle=(\cos
\theta- 1) \langle\xi, v_0-v
\rangle/2+ \sin\theta
\langle\xi, \Gamma(v_0-v,\varphi)  \rangle/2$. Hence
\begin{eqnarray*}
&&\int_0^{2\pi} \bigl(1-\cos\bigl(
\epsilon^{-1/\nu} \bigl\langle\xi, a(v_0,v,\theta,\varphi )
\bigr\rangle\bigr) \bigr)\,d\varphi
\\
&&\qquad=\int_0^{2\pi} \bigl(1-\cos\bigl(
\epsilon^{-1/\nu}(\cos\theta- 1) \langle \xi, v_0-v \rangle/2
\bigr)\\
&&\hspace*{73pt}{}\times\cos \bigl(\epsilon^{-1/\nu}\sin\theta \bigl\langle\xi, \Gamma
(v_0-v,\varphi) \bigr\rangle /2\bigr)
\\
&&\hspace*{35pt}\qquad{}+ \sin\bigl(\epsilon^{-1/\nu}(\cos\theta- 1) \langle\xi,
v_0-v \rangle /2\bigr)\\
&&\hspace*{51.5pt}\qquad{}\times\sin \bigl(\epsilon^{-1/\nu}\sin\theta
\bigl\langle\xi, \Gamma (v_0-v,\varphi) \bigr\rangle /2\bigr) \bigr)
\,d\varphi
\\
&&\qquad=\int_0^{2\pi} \bigl(1-\cos\bigl(
\epsilon^{-1/\nu}(\cos\theta- 1) \langle\xi, v_0-v \rangle/2
\bigr)\\
&&\hspace*{45pt}\qquad{}\times\cos \bigl(\epsilon^{-1/\nu}\sin\theta \bigl\langle\xi, \Gamma
(v_0-v,\varphi) \bigr\rangle /2\bigr) \bigr)\,d\varphi
\\
&&\qquad\geq \int_0^{2\pi} \bigl(1-\bigl\llvert \cos
\bigl( \epsilon^{-1/\nu}\sin\theta \bigl\langle\xi, \Gamma(v_0-v,
\varphi) \bigr\rangle/2 \bigr)\bigr\rrvert \bigr)\,d\varphi.
\end{eqnarray*}
Since $1-\cos x \geq x^2/4$ and $\llvert \sin x\rrvert  \geq\llvert x\rrvert /2$ for $x\in[-1,1]$
and since $\llvert \sin x\rrvert \leq\llvert x\rrvert $
for all $x \in\rr$ (recall that $\theta\leq\epsilon^{1/\nu}\leq1$),
\begin{eqnarray*}
&&\int_0^{2\pi} \bigl(1-\cos\bigl(
\epsilon^{-1/\nu} \bigl\langle\xi, a(v_0,v,\theta,\varphi )
\bigr\rangle\bigr) \bigr)\,d\varphi
\\
&&\qquad\geq \int_0^{2\pi} \frac{\epsilon^{-2/\nu}\sin^2 \theta
\langle\xi
, \Gamma
(v_0-v,\varphi)  \rangle^2}{16}
\indiq_{\{\llvert  \langle\xi, \Gamma(v_0-v,\varphi)  \rangle
\sin\theta\rrvert \leq2
\epsilon
^{1/\nu}\}}\, d\varphi
\\
&&\qquad\geq \int_0^{2\pi} \frac{\epsilon^{-2/\nu} \theta^2
\langle\xi,
\Gamma
(v_0-v,\varphi)  \rangle^2}{64}
\indiq_{\{\llvert \theta\rrvert  \leq2 \epsilon^{1/\nu}/\llvert  \langle\xi,
\Gamma
(v_0-v,\varphi)  \rangle
\rrvert  \}} \,d\varphi.
\end{eqnarray*}
Using the lower bound of $b$ given by (\ref{aaa}) and then
integrating in $\theta$, we obtain
\begin{eqnarray*}
&&\Re\Psi_{\epsilon,t,v_0} \bigl(\epsilon^{-1/\nu}\xi\bigr)
\\
&&\qquad\geq \frac{c}{\epsilon^{2/\nu}} \int_{t-\epsilon}^t \int
_{\rr^3}\int_0^{\epsilon^{1/\nu}} \int
_0^{2\pi
} \theta^2 \bigl\langle\xi ,
\Gamma(v_0-v,\varphi) \bigr\rangle^2
\\
&&\hspace*{136pt}{}\times\indiq_{\{\llvert \theta\rrvert  \leq2 \epsilon^{1/\nu}/\llvert  \langle\xi,
\Gamma
(v_0-v,\varphi)  \rangle
\rrvert  \}} \\
&&\hspace*{136pt}{}\times\llvert v-v_0\rrvert ^\gamma\, d
\varphi\theta^{-1-\nu} \,d\theta f_s(dv)\,ds
\\
&&\qquad= \frac{c}{\epsilon^{2/\nu}} \int_{t-\epsilon}^t \int
_{\rr^3}\int_0^{2\pi} \bigl
\langle\xi, \Gamma (v_0-v,\varphi) \bigr\rangle^2
\biggl[ \epsilon^{1/\nu}\land\frac{2\epsilon^{1/\nu}} {\llvert
\langle\xi,
\Gamma
(v_0-v,\varphi)  \rangle\rrvert } \biggr]^{2-\nu}\\
&&\hspace*{110pt}{}\times
\llvert v-v_0\rrvert ^\gamma f_s(dv)\,d\varphi
\,ds
\\
&&\qquad\geq \frac{c} \epsilon\int_{t-\epsilon}^t \int
_{\rr^3}\int_0^{2\pi} \bigl[
\bigl\langle\xi, \Gamma(v_0-v,\varphi) \bigr\rangle^2
\land\bigl\llvert \bigl\langle\xi, \Gamma(v_0-v,\varphi) \bigr
\rangle \bigr\rrvert ^{\nu} \bigr]\\
&&\hspace*{96pt}{}\times \llvert v-v_0\rrvert
^\gamma f_s(dv)\,d\varphi\, ds
\\
&&\qquad= \frac{c} \epsilon\int_{t-\epsilon}^t \int
_{\rr^3}\int_0^{2\pi} \bigl[
\bigl\langle v_0-v, \Gamma (\xi,\varphi) \bigr\rangle^2
\land\bigl\llvert \bigl\langle v_0-v, \Gamma(\xi,\varphi) \bigr
\rangle \bigr\rrvert ^{\nu} \bigr] \\
&&\hspace*{98pt}{}\times\llvert v-v_0\rrvert
^\gamma f_s(dv)\,d\varphi\, ds,
\end{eqnarray*}
where we finally used Remark~\ref{perp}.

\textit{Step 2.} We now assume that $\gamma\in(0,1)$.
Recall Proposition~\ref{lower2} [and the fact that $\llvert \Gamma(\xi
,\varphi
)\rrvert =\llvert \xi\rrvert $, see
(\ref{dfvprime})]: for any $v_0,\xi\in\rr^3$, any $\varphi\in
[0,2\pi)$,
any $v\in K(v_0,\Gamma(\xi,\varphi))$,
we have $\llvert v-v_0\rrvert \geq1$ and $\llvert  \langle v_0-v, \Gamma(\xi,\varphi
)  \rangle
\rrvert \geq
\llvert \Gamma(\xi,\varphi)\rrvert
=\llvert \xi\rrvert $. Thus, using that $f_s(K(v_0,\Gamma(\xi,\varphi)))\geq
q_{t_0,t_1}>0$ for all
$0< t_0\leq t-\epsilon\leq s \leq t \leq t_1$, we get
\begin{eqnarray*}
\Re\Psi_{\epsilon,t,v_0} \bigl(\epsilon^{-1/\nu}\xi\bigr) &\geq&
\frac{c} \epsilon\int_{t-\epsilon}^t \int
_0^{2\pi} \bigl[\llvert \xi\rrvert ^2
\land\llvert \xi\rrvert ^\nu \bigr] f_s\bigl(K
\bigl(v_0,\Gamma(\xi,\varphi)\bigr)\bigr)\,d\varphi \,ds \\&\geq& c
q_{t_0,t_1} \bigl[\llvert \xi\rrvert ^2 \land\llvert \xi\rrvert
^\nu \bigr].
\end{eqnarray*}

\textit{Step 3.} We finally assume that $\gamma\in(-1,0]$.
Recall again Proposition~\ref{lower2} and that $\llvert \Gamma(\xi,\varphi
)\rrvert =\llvert \xi\rrvert $:
for any $v_0,\xi\in\rr^3$, any $\varphi\in[0,2\pi)$,
any $v\in K(v_0,\break \Gamma(\xi,\varphi))$,
we have $\llvert v-v_0\rrvert \leq\llvert v\rrvert +\llvert v_0\rrvert  \leq3+\llvert v_0\rrvert $ [so that $\llvert v-v_0\rrvert ^\gamma
\geq3^\gamma(1+\llvert v_0\rrvert )^\gamma$]
and $\llvert  \langle v_0-v, \Gamma(\xi,\varphi)  \rangle\rrvert \geq
\llvert \Gamma(\xi,\varphi)\rrvert =
\llvert \xi\rrvert $.
Thus, using that $f_s(K(v_0,\break\Gamma(\xi,\varphi)))\geq q_{t_0,t_1}>0$
for all
$0< t_0\leq t-\epsilon\leq s \leq t \leq t_1$, we get
\begin{eqnarray*}
\Re\Psi_{\epsilon,t,v_0} \bigl(\epsilon^{-1/\nu}\xi\bigr)& \geq&
\frac{c} \epsilon\int_{t-\epsilon}^t \int
_0^{2\pi} \bigl[\llvert \xi\rrvert ^2
\land\llvert \xi\rrvert ^\nu \bigr] \bigl(1+\llvert v_0
\rrvert \bigr)^\gamma f_s\bigl(K\bigl(v_0,
\Gamma(\xi ,\varphi)\bigr)\bigr)\,d\varphi \,ds
\\
&\geq& c q_{t_0,t_1} \bigl(1+\llvert v_0\rrvert
\bigr)^\gamma \bigl[\llvert \xi\rrvert ^2 \land\llvert \xi
\rrvert ^\nu \bigr],
\end{eqnarray*}
which completes the proof.
\end{pf}

We now estimate the regularity of the law of $U^\epsilon_t$.

\begin{lem}\label{estibol2}
Adopt the notation and assumptions of Proposition~\ref{estibol3}.
Recall that
$\Psi_{\epsilon,t,v_0}$ was defined in Lemma~\ref{psibol}.
Consider $g_{\epsilon,t,v_0} \in\cP(\rr^3)$ such that $\widehat
{g_{\epsilon
,t,v_0}}(\xi)=\exp(-\Psi_{\epsilon,t,v_0}(\xi))$.
If $0<t_0\leq t-\epsilon<t \leq t_1$ and $\epsilon\in(0,1)$,
$g_{\epsilon,t,v_0}$ has a
density and
\[
\llVert \nabla g_{\epsilon,t,v_0}\rrVert _{L^1(\rr^3)} \leq
\cases{\displaystyle C_{t_0,t_1} \epsilon^{-1/\nu} \bigl(1+\llvert
v_0\rrvert \bigr)^{4\gamma+4} &\quad {if} $\gamma\in (0,1)$,
\cr\displaystyle
C_{t_0,t_1} \epsilon^{-1/\nu} \bigl(1+\llvert v_0
\rrvert \bigr)^{4+\gamma+4\llvert \gamma\rrvert /\nu} &\quad {if} $\gamma\in(-1,0]$. }
\]
\end{lem}

\begin{pf} We introduce, for $X_{\epsilon,t,v_0}$ a $g_{\epsilon
,t,v_0}$-distributed random variable,
$Y_{\epsilon,t,v_0}:=\epsilon^{-1/\nu}X_{\epsilon,t,v_0}$. Then the
law $k_{\epsilon,t,v_0}$ of
$Y_{\epsilon,t,v_0}$ satisfies
$\widehat{k_{\epsilon,t,v_0}}(\xi)=\break \widehat{g_{\epsilon
,t,v_0}}(\epsilon^{-1/\nu}\xi)=\exp
(-\Psi_{\epsilon,t,v_0}(\epsilon^{-1/\nu}\xi))$
and $k_{\epsilon,t,v_0}(x)=\epsilon^{3/\nu}g_{\epsilon
,t,v_0}(\epsilon^{1/\nu}x)$. Observe that
%
\begin{eqnarray}
\label{cqa} \llVert \nabla g_{\epsilon,t,v_0}\rrVert _{L^1(\rr^3)}=
\epsilon^{-1/\nu} \llVert \nabla k_{\epsilon
,t,v_0}\rrVert _{L^1(\rr^3)}.
\end{eqnarray}

\textit{Step 1.} We want to apply Lemma~\ref{ssw1}. We have
$\widehat{k_{\epsilon,t,v_0}}(\xi)=\exp(-\Phi_{\epsilon
,t,v_0}(\xi))$, where $\Phi
_{\epsilon,t,v_0}(\xi)=\Psi_{\epsilon,t,v_0}(\epsilon^{-1/\nu}\xi)$,
whence
\begin{eqnarray*}
\Phi_{\epsilon,t,v_0}(\xi)& =&\int_{t-\epsilon}^t \int
_{\rr^3}\int_0^{\epsilon^{1/\nu}} \int
_0^{2\pi} \bigl(1 - e^{i  \langle\xi, \epsilon^{-1/\nu
}a(v_0,v,\theta,\varphi
)  \rangle} \bigr)\\
&&\hspace*{82pt}{}\times
\llvert v-v_0\rrvert ^\gamma \,d\varphi b(\theta)\,d\theta
f_s(dv)\,ds
\\
&=&\int_{\rr^3}\bigl(1-e^{i  \langle\xi,z \rangle}\bigr) \lambda
_{t,\epsilon,v_0}(dz),
\end{eqnarray*}
the measure $\lambda_{t,\epsilon,v_0}$ being defined by
\begin{eqnarray*}
&&\int_{\rr^3}F(z) \lambda_{t,\epsilon,v_0}(dz)\\
&&\qquad=\int
_{t-\epsilon}^t \int_{\rr^3} \int
_0^{\epsilon
^{1/\nu}}\int_0^{2\pi}
F \biggl( \frac{a(v_0,v,\theta,\varphi)}{\epsilon^{1/\nu}} \biggr) \llvert v-v_0\rrvert
^\gamma \,d\varphi b(\theta)\,d\theta f_s(dv)\,ds %
\end{eqnarray*}
for all nonnegative measurable $F\dvtx  \rr^3 \mapsto\rr$. Lemma~\ref
{ssw1} thus implies
%
\begin{eqnarray}
\label{sss} \llVert \nabla k_{\epsilon,t,v_0}\rrVert _{L^1(\rr^3)} &\leq& C
\bigl(1+m_1^4(\lambda_{t,\epsilon
,v_0})+m_4 (
\lambda_{t,\epsilon,v_0}) \bigr) \nonumber\\
&&{}\times\int_{\rr^3}e^{-\Re\Phi_{\epsilon,t,v_0}(\xi)}
\bigl(1+\llvert \xi\rrvert \bigr)\,d\xi
\nonumber
\\[-8pt]\\[-8pt]
&\leq& C \bigl(1+m_1^4(\lambda_{t,\epsilon,v_0})+m_4
(\lambda _{t,\epsilon,v_0}) \bigr)\nonumber\\
&&{}\times \biggl( 1+ \int_{\llvert \xi\rrvert \geq1}
e^{-\Re\Psi_{\epsilon,t,v_0}(\epsilon^{-1/\nu
}\xi)} \llvert \xi\rrvert \,d\xi \biggr).\nonumber
\end{eqnarray}
A simple computation using (\ref{lit}) and
(\ref{aaa}) shows that for $n=1,4$,
%
\begin{eqnarray}
\label{mll} m_n(\lambda_{t,\epsilon,v_0}) &\leq& \int
_{t-\epsilon}^t \int_{\rr^3} \int
_0^{\epsilon^{1/\nu
}}\int_0^{2\pi}
\frac{\llvert \theta\rrvert ^n\llvert v-v_0\rrvert ^n}{2^n\epsilon^{n/\nu}} \llvert v-v_0\rrvert ^\gamma\, d\varphi
b(\theta)\,d\theta f_s(dv)\,ds\nonumber
\\
&\leq&C \int_{t-\epsilon}^t \int_{\rr^3}
\int_0^{\epsilon^{1/\nu}} \frac{\llvert \theta\rrvert ^{n-1-\nu}\llvert v-v_0\rrvert ^{n+\gamma}}{\epsilon^{n/\nu}} \,d\theta
f_s(dv)\,ds
\nonumber
\\[-8pt]\\[-8pt]
&\leq& C \int_{t-\epsilon}^t \int_{\rr^3}
\bigl(\llvert v\rrvert ^{\gamma
+n}+\llvert v_0\rrvert
^{\gamma+n}\bigr) \frac{\epsilon^{(n-\nu)/\nu}}{\epsilon^{n/\nu}} f_s(dv)\,ds
\nonumber
\\
&\leq& C \sup_{s\in[t-\epsilon,t]} \int_{\rr^3}\bigl(
\llvert v\rrvert ^{\gamma
+n}+\llvert v_0\rrvert
^{\gamma
+n}\bigr)f_s(dv).
\nonumber
\end{eqnarray}

\textit{Step 2.} Here we conclude when $\gamma\in(0,1)$. Let thus
$0<t_0\leq t-\epsilon\leq t \leq t_1$ with
$\epsilon\in(0,1)$. Using (\ref{energy}), we deduce that
$\sup_{s\in[t-\epsilon,t]} \int_{\rr^3}(\llvert v\rrvert ^{\gamma
+1}+\llvert v_0\rrvert ^{\gamma+1})f_s(dv)
\leq C(1+\llvert v_0\rrvert ^{\gamma+1})$
and by (\ref{mom}), $\sup_{s\in[t-\epsilon,t]} \int_{\rr
^3}(\llvert v\rrvert ^{\gamma
+4}+\llvert v_0\rrvert ^{\gamma+4})\*f_s(dv) \leq
C_{t_0}(1+\llvert v_0\rrvert ^{\gamma+4})$. Hence $m_1^4(\lambda_{t,\epsilon
,v_0})+m_4(\lambda_{t,\epsilon,v_0})\leq
C_{t_0}(1+\llvert v_0\rrvert ^{4\gamma+4})$.
By Lemma~\ref{estibol},
$\int_{\llvert \xi\rrvert \geq1} e^{-\Re\Psi_{\epsilon,t,v_0}(\epsilon^{-1/\nu
}\xi)} \llvert \xi\rrvert \,d\xi
\leq C_{t_0,t_1}$. Recalling
(\ref{sss}), we finally find that $\llVert \nabla k_{\epsilon
,t,v_0}\rrVert _{L^1(\rr^3)}
\leq
C_{t_0,t_1}(1+\llvert v_0\rrvert ^{4\gamma+4})$, whence the result by (\ref{cqa}).

\textit{Step 3.} We finally conclude when $\gamma\in(-1,0]$. Let thus
$0<t_0\leq t-\epsilon\leq t \leq t_1$ with
$\epsilon\in(0,1)$. Using (\ref{energy}), we deduce that
$\sup_{s\in[t-\epsilon,t]} \int_{\rr^3}(\llvert v\rrvert ^{\gamma
+1}+\llvert v_0\rrvert ^{\gamma+1})f_s(dv)
\leq C(1+\llvert v_0\rrvert ^{\gamma+1})$.
By (\ref{mom2}) and since $f_0 \in\cP_{4+\gamma+4\llvert \gamma\rrvert /\nu
}(\rr^3)
\subset\cP_{4+\gamma}(\rr^3)$,
we deduce that $\sup_{s\in[t-\epsilon,t]} \int_{\rr^3}(\llvert v\rrvert ^{\gamma
+4}+\llvert v_0\rrvert ^{\gamma
+4})f_s(dv) \leq
C_{t_1}(1+\llvert v_0\rrvert ^{\gamma+4})$. Hence $m_1^4(\lambda_{t,\epsilon
,v_0})+m_4(\lambda_{t,\epsilon,v_0})\leq
C_{t_1}(1+\llvert v_0\rrvert ^{\gamma+4})$.
By Lemma~\ref{estibol},
$\int_{\llvert \xi\rrvert \geq1} e^{-\Re\Psi_{\epsilon,t,v_0}(\epsilon^{-1/\nu
}\xi)} \llvert \xi\rrvert \,d\xi
\leq
\int_{\llvert \xi\rrvert \geq1} e^{-c_{t_0,t_1} (1+\llvert v_0\rrvert )^\gamma\llvert
 \xi\rrvert ^\nu}\llvert \xi\rrvert \,d\xi\leq C_{t_0,t_1} (1+\break \llvert v_0\rrvert )^{4\llvert \gamma\rrvert /\nu}$.
Recalling (\ref{sss}), we finally get $\llVert \nabla k_{\epsilon
,t,v_0}\rrVert _{L^1(\rr
^3)} \leq
C_{t_0,t_1}(1+\llvert v_0\rrvert ^{\gamma+4}) \*(1+\llvert v_0\rrvert )^{4\llvert \gamma\rrvert /\nu}\leq
C_{t_0,t_1}(1+\llvert v_0\rrvert )^{4+\gamma+4\llvert \gamma\rrvert /\nu}$, whence the result by
(\ref{cqa}).
\end{pf}

We finally have all the weapons to give the following:

\begin{pf*}{Proof of Lemma~\ref{estibol3}}
Let thus $t_0\leq t-\epsilon\leq t\leq t_1$ with $\epsilon\in(0,1)$,
and let $\phi
\in L^\infty(\rr^3)$.
Recall the notation introduced in Lemma~\ref{psibol}. Write, using
that $W^\epsilon_t$ and $U^\epsilon_t$ are independent conditionally
on $\cF_{t-\epsilon}$
and that the law of $U^\epsilon_t$ conditionally on $\cF_{t-\epsilon
}$ is $g_{\epsilon
,t,V_{t-\epsilon}}$ (see Lemma~\ref{estibol2})
\begin{eqnarray*}
&&\bigl\llvert \E\bigl[\phi\bigl(V^\epsilon_t+h\bigr)-\phi
\bigl(V^\epsilon_t\bigr)\bigr]\bigr\rrvert \\
&&\qquad=\bigl\llvert \E
\bigl[\phi \bigl(U^\epsilon_t+W^\epsilon_t+h
\bigr)-\phi\bigl(U^\epsilon _t+W^\epsilon_t
\bigr)\bigr]\bigr\rrvert
\\
&&\qquad=\bigl\llvert \E \bigl[ \E \bigl( \phi\bigl(U^\epsilon_t+W^\epsilon_t+h
\bigr)-\phi \bigl(U^\epsilon_t+W^\epsilon_t
\bigr) \vert \cF_{t-\epsilon} \bigr) \bigr]\bigr\rrvert
\\
&&\qquad= \biggl\llvert \E \biggl[\int_{\rr^3}\bigl[\phi
\bigl(x+W^\epsilon_t+h\bigr) - \phi \bigl(x+W^\epsilon_t
\bigr)\bigr] g_{\epsilon
,t,V_{t-\epsilon}}(x)\,dx \biggr] \biggr\rrvert
\\
&&\qquad= \biggl\llvert \E \biggl[ \int_{\rr^3}\phi
\bigl(x+W^\epsilon_t\bigr)\bigl[g_{\epsilon
,t,V_{t-\epsilon}}(x-h) -
g_{\epsilon
,t,V_{t-\epsilon}}(x)\bigr]\,dx \biggr] \biggr\rrvert
\\
&&\qquad\leq \llVert \phi\rrVert _{L^\infty(\rr^3)} \llvert h\rrvert \E \bigl[\llVert
\nabla g_{\epsilon
,t,V_{t-\epsilon
}}\rrVert _{L^1(\rr^3)} \bigr].
\end{eqnarray*}
We used that $\int_{\rr^3}\llvert g(x-h)-g(x)\rrvert \,dx \leq\int_{\rr^3}\int_0^1 \llvert h.\nabla
g(x-uh)\rrvert \,du\, dx\leq
\llvert h\rrvert\* \int_0^1 \llVert \nabla g(\cdot -uh) \rrVert _{L^1(\rr^3)}\, du=\llvert h\rrvert \llVert \nabla
g\rrVert _{L^1(\rr^3)}$.

Assume first that $\gamma\in(0,1)$. Using Lemma~\ref{estibol2}, we get
\begin{eqnarray*}
\bigl\llvert \E\bigl[\phi\bigl(V^\epsilon_t+h\bigr)-\phi
\bigl(V^\epsilon_t\bigr)\bigr]\bigr\rrvert \leq
C_{t_0,t_1} \llVert \phi\rrVert _{L^\infty
(\rr^3)} \llvert h\rrvert
\epsilon^{-1/\nu} \E \bigl[\bigl(1+\llvert V_{t-\epsilon}\rrvert
\bigr)^{4\gamma+4} \bigr].
\end{eqnarray*}
The conclusion follows, since
\[
\E\bigl [\llvert V_{t-\epsilon}\rrvert ^{4\gamma+4}\bigr]=m_{4\gamma+4}(f_{t-\epsilon})\leq
\sup_{s\geq t_0} m_{4\gamma+4}(f_s)<\infty
\]
by (\ref{mom}).

Assume next that $\gamma\in(-1,0]$. In this case, Lemma~\ref
{estibol2} gives
\begin{eqnarray*}
\bigl\llvert \E\bigl[\phi\bigl(V^\epsilon_t+h\bigr)-\phi
\bigl(V^\epsilon_t\bigr)\bigr]\bigr\rrvert \leq
C_{t_0,t_1} \llVert \phi\rrVert _{L^\infty
(\rr^3)} \llvert h\rrvert
\epsilon^{-1/\nu} \E \bigl[\bigl(1+\llvert V_{t-\epsilon}\rrvert
\bigr)^{4+\gamma+4\llvert \gamma\rrvert /\nu} \bigr].
\end{eqnarray*}
But since $f_0 \in\cP_{4+\gamma+4\llvert \gamma\rrvert /\nu}(\rr^3)$ and $0\leq
t-\epsilon
\leq t_1$,
(\ref{mom2}) implies that\break
$\E [\llvert V_{t-\epsilon}\rrvert ^{4+\gamma+4\llvert \gamma\rrvert /\nu}
]=m_{4+\gamma
+4\llvert \gamma\rrvert /\nu}(f_{t-\epsilon})\leq C_{t_1}$,
which completes the proof.
\end{pf*}

\section{Conclusion}\label{conclu}

We finally can give the following:

\begin{pf*}{Proof of Theorem~\ref{mr}}
We thus assume (\ref{aaa}) for some $\gamma\in(-1,1)$, $\nu
\in(0,1)$ such that
$\gamma+\nu>0$. We also consider $f_0 \in\cP_2(\rr^3)$ such that $f_0$
is not a Dirac mass.
If $\gamma\in(0,1)$, we consider any weak solution
$(f_t)_{t\geq0}$ to (\ref{be}) starting from $f_0$ and satisfying
(\ref
{mom}) and we consider
the associated Boltzmann process $(V_t)_{t\geq0}$ built in Proposition~\ref{exifc}(ii).
If $\gamma\in(-1,0]$, we assume additionally that $f_0 \in\cP
_{4+\gamma+4\llvert \gamma\rrvert /\nu}(\rr^3)$, and
we consider the weak solution
$(f_t)_{t\geq0}$ to (\ref{be}) starting from $f_0$ and
the associated Boltzmann process $(V_t)_{t\geq0}$ built in Proposition~\ref{exifc}(ii).
From now on, we fix $t>0$.

We wish to apply Lemma~\ref{thelem}. Let thus $h\in\rr^3$ such that
$\llvert h\rrvert \leq1$ and
$\phi\in C^\alpha_b(\rr^3)$ for some $\alpha\in(0,1)$. Let us define
\[
I_{t,h}^\phi= \biggl\llvert \int_{\rr^3}
\bigl(\phi(v+h)-\phi(v)\bigr)f_t(dv) \biggr\rrvert = \bigl\llvert \E
\bigl[\phi(V_t+h)-\phi(V_t) \bigr] \bigr\rrvert .
\]
For $\epsilon\in(0,(t/2)\land1)$, we write, recalling that the approximate
Boltzmann process
$V_t^\epsilon$ was defined in Lemma~\ref{app},
\begin{eqnarray*}
I_{t,h}^\phi&\leq& \bigl\llvert \E\bigl[
\phi(V_t+h)-\phi\bigl(V_t^\epsilon+h\bigr)\bigr]
\bigr\rrvert + \bigl\llvert \E \bigl[\phi(V_t)-\phi
\bigl(V_t^\epsilon\bigr)\bigr]\bigr\rrvert \\
&&{}+ \bigl\llvert \E
\bigl[\phi\bigl(V_t^\epsilon+h\bigr)-\phi\bigl(V_t^\epsilon
\bigr)\bigr]\bigr\rrvert
\\
&\leq& 2 \llVert \phi\rrVert _{C^\alpha_b(\rr^3)}\E\bigl[\bigl\llvert
V_t-V^\epsilon_t\bigr\rrvert ^\alpha
\bigr]+ C_t \llVert \phi \rrVert _\infty\epsilon^{-1/\nu}
\llvert h\rrvert
\\
&\leq& C_t \llVert \phi\rrVert _{C^\alpha_b(\rr^3)} \bigl[\E\bigl[\bigl
\llvert V_t-V^\epsilon _t\bigr\rrvert
^\alpha\bigr] + \epsilon^{-1/\nu}\llvert h\rrvert \bigr],
\end{eqnarray*}
where we used Lemma~\ref{estibol3} (with $t_0=t/2$ and $t_1=t$) and
that $\llVert \phi\rrVert _{L^\infty(\rr^3)}\leq
\llVert \phi\rrVert _{C^\alpha_b(\rr^3)}$.

\textit{Point} (i). We assume here that $\gamma\in(0,1)$. We consider
$\alpha\in(0,\nu]$, and we apply Proposition~\ref{app}(i): for any $\eta\in(0,2)$, we write
$\E[\llvert V_t-V^\epsilon_t\rrvert ^\alpha] \leq\E[\llvert V_t-V^\epsilon_t\rrvert ^{\nu
}]^{\alpha/\nu} \leq
C_{t,\eta} \epsilon^{(2-\eta)\alpha/\nu}$.
We have proved that
for all $\eta\in(0,2)$, all $\epsilon\in(0,(t/2)\land1)$,
\[
I_{t,h}^\phi\leq C_{t,\eta} \llVert \phi\rrVert
_{C^\alpha_b(\rr^3)} \bigl[\epsilon ^{(2-\eta) \alpha/\nu} + \epsilon^{-1/\nu}\llvert
h\rrvert \bigr]. %
\]
Choosing $\epsilon=(1\land(t/2))\llvert h\rrvert ^{\nu/(1+(2-\eta)\alpha)}$, we obtain
$I_{t,h}^\phi\leq C_{t,\eta} \llVert \phi\rrVert _{C^\alpha_b(\rr^3)}\times\break
\llvert h\rrvert ^{\fracc
{(2-\eta)\alpha}{1+(2-\eta)\alpha}}$.
For $\alpha\in(0,\nu]$ small enough and $\eta\in(0,2)$ small
enough, it
holds that
$\frac{(2-\eta)\alpha}{1+(2-\eta)\alpha}>\alpha$.
Applying Lemma~\ref{thelem}, we deduce that $f_t$ has a density with
furthermore
$f_t \in B^s_{1,\infty}(\rr^3)$ for any $s\in(0,s_\nu)$, where
\begin{eqnarray*}
s_\nu=\sup \biggl\{\frac{(2-\eta)\alpha}{1+(2-\eta)\alpha
}-\alpha \dvtx  \alpha\in(0,\nu],
\eta\in(0,2) \biggr\}.
\end{eqnarray*}
It is easily checked that $s_\nu$ is given by (\ref{snu}).

\textit{Point} (ii). We next assume that $\gamma\in(-1,0]$ and that
$\gamma+\nu>0$. We consider
$\alpha\in(0,\nu]$ and we apply Proposition~\ref{app}(ii): for any $\eta\in(0,2+\gamma/\nu)$,
$\E[\llvert V_t-V^\epsilon_t\rrvert ^\alpha] \leq\E[\llvert V_t-V^\epsilon_t\rrvert ^{\nu
}]^{\alpha/\nu} \leq
C_{t,\eta} \epsilon^{(2+\gamma/\nu-\eta)\alpha/\nu}$.
Hence
for all $\eta\in(0,2+\gamma/\nu)$, all $\epsilon\in(0,(t/2)\land1)$,
\[
I_{t,h}^\phi\leq C_{t,\eta} \llVert \phi\rrVert
_{C^\alpha_b(\rr^3)} \bigl[\epsilon ^{(2+\gamma/\nu-\eta)\alpha/\nu} + \epsilon^{-1/\nu}\llvert
h\rrvert \bigr]. %
\]
Choosing $\epsilon=(1\land(t/2))\llvert h\rrvert ^{\nu/(1+(2+\gamma/\nu-\eta
)\alpha)}$, we obtain
$I_{t,h}^\phi\leq C_{t,\eta} \llVert \phi\rrVert _{C^\alpha_b(\rr^3)}\*
\llvert h\rrvert ^{\fracc
{(2+\gamma/\nu-\eta)\alpha}{1+(2+\gamma/\nu-\eta)\alpha}}$.
For $\alpha\in(0,\nu]$ small enough and $\eta\in(0,2+2\gamma/\nu)$
small enough, it holds that
$\frac{(2+\gamma/\nu-\eta)\alpha}{1+(2+\gamma/\nu-\eta)\alpha
}>\alpha$
(because $2+\gamma/\nu>1$).
Applying Lemma~\ref{thelem}, we deduce that $f_t$ has a density with
furthermore
$f_t \in B^s_{1,\infty}(\rr^3)$ for any $s\in(0,s_{\gamma,\nu})$, where
\begin{eqnarray*}
s_{\gamma,\nu}=\sup \biggl\{\frac{(2+\gamma/\nu-\eta)\alpha
}{1+(2+\gamma
/\nu-\eta)\alpha}-\alpha \dvtx  \alpha\in(0,\nu],
\eta\in(0,2+\gamma/\nu) \biggr\}.
\end{eqnarray*}
It is easily checked that $s_{\gamma,\nu}$ is given by (\ref{sganu}).

\textit{Point} (iii). In any case, we thus have $f_t \in B^s_{1,\infty
}(\rr
^3)$ for some
$s>0$. This implies that $f_t \in L^p(\rr^3)$ for all $p \in
(1,3/(3-s))$; see, for example,
\cite{rs}, Corollary~2(ii), page 36.
The facts that $f_t \in\cP_2(\rr^3) \cap L^p(\rr^3)$ for some $p>1$
classically imply that $\int_{\rr^3}f_t(v)\llvert \log f_t(v)\rrvert \,dv <\infty$.
\end{pf*}

\section{Existence of the Boltzmann process}\label{exifcs}

It remains to prove Proposition~\ref{exifc}. We have already checked
very similar results in
several closely related situations, but always with some restrictions
(in the $2D$-case or for
bounded velocity cross sections or assuming conditions on the initial
data that guarantees uniqueness
of the solution). We thus give a rather complete proof. Unfortunately,
we have to treat separately the case of hard and moderately soft potentials:
for hard potentials, we associate a Boltzmann process to any weak solution,
while for moderately soft potentials, we can only build one Boltzmann
process, which corresponds
to one weak solution. Thus the proofs really differ.

\subsection{Moderately soft potentials}

In the whole subsection, we assume (\ref{aaa})
for some $\gamma\in(-1,0]$, $\nu\in(0,1)$, and we consider
$f_0\in\cP_p(\rr^3)$ for some $p>2$.
We want to prove Proposition~\ref{exifc}(ii).
Recall that $L_B$ was
defined in (\ref{dfL}) and rewritten in (\ref{newL}).

\begin{defi} Let $B(\llvert v-v_*\rrvert ,\cos\theta)$ be a given cross section.
A c\`adl\`ag adapted process $(V_t)_{t\geq0}$ on some probability space
$(\Omega,\cF,(\cF_t)_{t\geq0},\Pr)$ is said to solve
the martingale problem $\MP(f_0,B)$ if:
\begin{longlist}[(b)]
\item[(a)] $\cL(V_0)=f_0$,

\item[(b)] for all $t\geq0$, $\E[V_t]=\int_{\rr^3}v f_0(dv)$ and $\E
[\llvert V_t\rrvert ^2]=\int_{\rr^3}\llvert v\rrvert ^2f_0(dv)$,

\item[(c)] for all $\phi\in\lip_b(\rr^3)$, $(M_t^\phi)_{t\geq0}$
is a $(\Omega,\cF,(\cF_t)_{t\geq0},\Pr)$-martingale, where
$M^\phi_t:=\phi(V_t)-\int_0^t \int_{\rr^3}L_B\phi(V_s,v)
f_s(dv)\,ds$ and where
$f_t:=\cL(V_t)$.
\end{longlist}
\end{defi}

The following remarks are classical.

\begin{rk}\label{eq}
(i) A c\`adl\`ag adapted process $(V_t)_{t\geq0}$ on some probability
space $(\Omega,\cF,(\cF_t)_{t\geq0},\Pr)$
is a solution to $\MP(f_0,B)$ if and only if it satisfies point (a) and
(b) of the above
definition and if
there exists, on a possibly enlarged probability space, a $(\cF
_t)_{t\geq0}$-Poisson measure
$N(ds,dv,d\theta,d\varphi,du)$ on
$[0,\infty)\times\rr^3 \times(0,\pi/2]\times[0,2\pi)\times
[0,\infty)$
with intensity $dsf_s(dv) b(\theta)\,d\theta \,d\varphi \,du$ [where
$f_t:=\cL(V_t)$]
such that $(V_t)_{t\geq0}$ solves (\ref{bolsde}).

(ii) If $(V_t)_{t\geq0}$ solves $\MP(f_0,B)$ and if $f_t:=\cL(V_t)$,
then $(f_t)_{t\geq0}$ is a weak solution to (\ref{be}) starting from $f_0$.
\end{rk}

See, for example, Tanaka \cite{t}, Section~4, for (i). Point (ii) is
obvious: use that for
$\phi\in\lip_b(\rr^3)$, for $t\geq0$, $\E[M^\phi_t]=\E[M^\phi
_0]=\E
[\phi(V_0)]$.

We start with the following statement.

\begin{rk}\label{eub} Let $B$ be a cross section satisfying
(\ref{aaa})
for some $\gamma\in(-1,0]$, $\nu\in(0,1)$. For $k\geq1$, define
$B_{k}(\llvert v-v_*\rrvert ,\cos\theta)\sin\theta=(\llvert v-v_*\rrvert ^\gamma\land k)
b(\theta
) \indiq_{\{\theta>1/k\}}$.
There exists a (unique in law) solution to $(V^{k}_t)_{t\geq0}$ to\break
$\MP(f_0,B_{k})$.
\end{rk}

This result can be checked easily, because $\int_0^{\pi/2} b(\theta
)\indiq_{\{\theta>1/k\}}\,d\theta<\infty$
and because $(\llvert z\rrvert ^\gamma\land k)$ is bounded. For example, one can use
a perfect simulation algorithm, see, for example, \cite{fg} for a very
similar result concerning the
Smoluchowski equation.

Below, $\dd([0,\infty),\rr^3)$ stands for the set of $\rr^3$-valued
c\`
adl\`ag functions,
which we endow with the Skorokhod topology; see, for example, Jacod and
Shiryaev \cite{js}.

\begin{lem}\label{tight}
Adopt the assumptions and notation of Remark~\ref{eub} and recall that
$f_0 \in\cP_p(\rr^3)$
for some $p>2$:
\begin{longlist}[(iii)]
\item[(i)] for all $T>0$, $\sup_{k\geq1}\E[\sup_{[0,T]} \llvert V^k_t\rrvert ^p]\leq C_{T,p}$;

\item[(ii)] the family $((V^{k}_t)_{t\geq0})_{k\geq1}$ is tight in $\dd
([0,\infty),\rr^3)$ and any limit process
$(V_t)_{t\geq0}$ satisfies $\Pr(V_t\ne V_{t-})=0$ for all $t\geq0$;

\item[(iii)] any limit $(V_t)_{t\geq0}$ solves $\MP(f_0,B)$ and verifies
$\E[\sup_{[0,T]} \llvert V_t\rrvert ^p]\leq C_{T,p}$ for all $T>0$.
\end{longlist}
\end{lem}

\begin{pf} We start with (i).
Set $f^k_t:=\cL(V^k_t)$.
As in Remark~\ref{eq}(i), there is a Poisson measure
$N_k(ds,dv,d\theta,d\varphi,du)$ on $[0,\infty)\times\rr^3 \times
(0,\pi/2]\times[0,2\pi)\times[0,\infty)$
with intensity $dsf_s^k(dv) b(\theta)\,d\theta \,d\varphi\, du$ such that
\begin{eqnarray*}
&&V_t^k=V_0^k + \int
_0^t \int_{\rr^3}\int
_0^{\pi/2}\int_0^{2\pi
}
\int_0^\infty a\bigl(V_\sm^k,v,
\theta,\varphi\bigr)\\
&&\hspace*{150pt}{}\times\indiq_{\{u\leq\llvert V_\sm^k-v\rrvert ^\gamma
\land k\}
}\\
&&\hspace*{150pt}{}\times\indiq_{\{\theta>1/k\}}N_k(ds,dv,d
\theta,d\varphi,du).
\end{eqnarray*}
Observe now that due to (\ref{lit}),
\begin{eqnarray*}
&&\bigl\llvert \bigl\llvert V_\sm^k+a\bigl(V_\sm^k,v,
\theta,\varphi\bigr)\bigl\llvert ^p - \bigr\rrvert
V_\sm^k\bigl\llvert ^p \bigr\rrvert \\
&&\qquad\leq
C_p \bigl( \bigl\llvert V_\sm^k\bigr\rrvert
^{p-1} + \bigl\llvert a\bigl(V_\sm^k,v,\theta,
\varphi\bigr)\bigr\rrvert ^{p-1}\bigr) \bigl\llvert a
\bigl(V_\sm^k,v,\theta,\varphi\bigr)\bigr\rrvert
\\
&&\qquad\leq C_p \bigl(1+\bigl\llvert V_\sm^k\bigr
\rrvert ^{p-1}+\llvert v\rrvert ^{p-1}\bigr) \bigl\llvert
V_\sm^k-v\bigr\rrvert \theta
\end{eqnarray*}
so that, using the It\^o formula for jump process (see, e.g., Jacod and
Shiryaev \cite{js}, Theorem~4.57, page 56),
\begin{eqnarray*}
\hspace*{-3pt}&&\sup_{[0,t]}\bigl\llvert V_r^k\bigr
\rrvert ^p\\
\hspace*{-3pt}&&\qquad\leq\bigl\llvert V_0^k\bigr
\rrvert ^p + C_p \int_0^t
\int_{\rr^3}\int_0^{\pi
/2}\int
_0^{2\pi}\int_0^\infty
\bigl(1+\bigl\llvert V_\sm^k\bigr\rrvert
^{p-1}+\llvert v\rrvert ^{p-1}\bigr) \bigl\llvert
V_\sm^k-v\bigr\rrvert \theta
\\
\hspace*{-3pt}&&\hspace*{159pt}\qquad{}\times
\indiq_{\{u\leq\llvert V_\sm^k-v\rrvert ^\gamma\}}N_k(ds,dv,d\theta,d\varphi,du).
\end{eqnarray*}
Taking expectations and using that $\int_0^{\pi/2} \theta b(\theta
)\,d\theta<\infty$ by (\ref{aaa}), we get
\begin{eqnarray*}
\E \Bigl(\sup_{[0,t]}\bigl\llvert V_r^k
\bigr\rrvert ^p \Bigr)&\leq& \E\bigl(\bigl\llvert V_0^k
\bigr\rrvert ^p\bigr) \\
&&{}+ C_p \int_0^t
\int_{\rr^3} \E \bigl[\bigl(1+\bigl\llvert V_s^k
\bigr\rrvert ^{p-1}+\llvert v\rrvert ^{p-1}\bigr) \bigl\llvert
V_s^k-v\bigr\rrvert ^{1+\gamma} \bigr]
f_s^k(dv)\,ds.
\end{eqnarray*}
Since $\gamma+1 \in(0,1]$ and $f^k_t=\cL(V^k_t)$,
\begin{eqnarray*}
\E \Bigl(\sup_{[0,t]}\bigl\llvert V_r^k
\bigr\rrvert ^p \Bigr)&\leq& \E\bigl(\bigl\llvert V_0^k
\bigr\rrvert ^p\bigr) + C_p \int_0^t
\int_{\rr^3} \E \bigl[1+\bigl\llvert V_s^k
\bigr\rrvert ^{p}+\llvert v\rrvert ^{p} \bigr]
f_s^k(dv)\,ds
\\
&\leq& \E\bigl(\bigl\llvert V_0^k\bigr\rrvert
^p\bigr) + C_p \int_0^t
\E \bigl[1+\bigl\llvert V_s^k\bigr\rrvert
^{p} \bigr]\,ds.
\end{eqnarray*}
Finally, $\E(\llvert V_0^k\rrvert ^p)=m_p(f_0)<\infty$ does not depend on $k$ and
we conclude with the Gr\"onwall lemma.

To check (ii), we use the Aldous \cite{al} criterion (which shows both
tightness and that any limit
process has no fixed discontinuity); see also \cite{js}, page 321.
Due to (i), it suffices that
for all $T>0$,
%
\begin{eqnarray}
\label{cqdf} \lim_{\delta\to0} \sup_{k\geq1} \sup
_{(S,S') \in\cS_T(\delta)} \E \bigl[\bigl\llvert V^k_{S'}-V^k_S
\bigr\rrvert \bigr]=0,
\end{eqnarray}
the set $\cS_T(\delta)$ consisting of all pairs $(S,S')$ of stopping
times satisfying
$0\leq S \leq S'\leq S+\delta\leq T$. Let thus $T>0$, $\delta>0$,
$(S,S') \in\cS_T(\delta)$ and $k\geq1$ be fixed. Using the s.d.e.
satisfied by
$(V^k_t)_{t\geq0}$, we immediately get
\begin{eqnarray*}
\hspace*{-4pt}&&\E\bigl[\bigl\llvert V^k_{S'}-V^k_S
\bigr\rrvert \bigr] \\
\hspace*{-4pt}&&\hspace*{21pt}\leq\E \biggl[\int_S^{S+\delta}
\int_{\rr
^3}\int_0^{\pi
/2}\int
_0^{2\pi} \bigl\llvert a(V_s,v,\theta,
\varphi)\bigr\rrvert \bigl\llvert V^k_s-v\bigr\rrvert
^\gamma \,d\varphi b(\theta)\,d\theta\, d\varphi f_s^k(dv)
\,ds \biggr].
\end{eqnarray*}
Using (\ref{lit}), that $\int_0^{\pi/2}\theta b(\theta)\,d\theta
<\infty$
by (\ref{aaa})
and that $\int_{\rr^3}\llvert v\rrvert ^{\gamma+1} f^k_s(dv)= \E[\llvert V^k_s\rrvert ^{\gamma+1}]$
is bounded for $s\in[0,T]$ due to (i), this gives
\begin{eqnarray*}
\E\bigl[\bigl\llvert V^k_{S'}-V^k_S
\bigr\rrvert \bigr] &\leq& C\E \biggl[\int_S^{S+\delta}
\int_{\rr^3} \bigl\llvert V^k_s-v\bigr
\rrvert ^{\gamma+1} f_s^k(dv)\,ds \biggr] \\&\leq&
C_T \E \biggl[\int_S^{S+\delta}
\bigl(1+\bigl\llvert V^k_s\bigr\rrvert
\bigr)^{\gamma+1}\,ds \biggr].
\end{eqnarray*}
Finally,
\begin{eqnarray*}
\E\bigl[\bigl\llvert V^k_{S'}-V^k_S
\bigr\rrvert \bigr] \leq C_T\E \Bigl[\delta\sup_{[0,T]}
\bigl(1+\bigl\llvert V^k_s\bigr\rrvert
\bigr)^{\gamma+1} \Bigr] \leq C_T \delta
\end{eqnarray*}
by point (i), whence (\ref{cqdf}).

We finally check (iii). Let thus $(V_t)_{t\geq0}$ be the limit in law
of a (not relabelled)
subsequence of $(V_t^k)_{t\geq0}$.
Write $f_t:= \cL(V_t)$ and $f_t^k:= \cL(V_t^k)$.
First, we obviously have $\cL(V_0)=f_0$, since
$\cL(V_0^k)=f_0$ for all $k\geq1$.
We also have $\E[\sup_{[0,T]} \llvert V_t\rrvert ^p]\leq C_{T,p}$ for all $T>0$
thanks to point (i). Since
we have $\E[V_t^k]=\int_{\rr^3}vf_0(dv)$ and $\E[\llvert V_t^k\rrvert ^2]=\int_{\rr^3}
\llvert v\rrvert ^2f_0(dv)$ for all $k\geq1$ and all
$t\geq0$, we easily deduce from (i) (recall that $p>2$) that
$\E[V_t]=\int_{\rr^3}vf_0(dv)$ and $\E[\llvert V_t\rrvert ^2]=\int_{\rr
^3}\llvert v\rrvert ^2f_0(dv)$ for all
$t\geq0$.
It only remains to check that for all $\phi\in\lip_b(\rr^3)$,
$(M_t^\phi)_{t\geq0}$
is a martingale, where $M^\phi_t:=\phi(V_t)-\int_0^t \int_{\rr
^3}L_B\phi
(V_s,v)\* f_s(dv)\,ds$.
To do so, consider $n\geq1$, $0\leq t_1\leq\cdots\leq t_n \leq s \leq
t$ and a family of
continuous bounded functions $\phi_1,\dots,\phi_n$ on $\rr^3$. We have
to prove that
$\E[ \Psi_{B,f}(V)]=0$,
where, for $x\in\dd([0,\infty),\rr^3)$,
\[
\Psi_{B,f}(x)=\prod_{i=1}^n
\phi_i(x_{t_i}) \biggl(\phi(x_t)-
\phi(x_s)- \int_s^t \int
_{\rr^3}L_B\phi(x_r,v)
f_r(dv)\,dr \biggr).\vadjust{\goodbreak} %
\]
Since $(V_t^k)_{t\geq0}$ solves $\MP(f_0,B_k)$, we know that $\E[\Psi
_{B_k,f^k}(V^k)]=0$, where
$\Psi_{B_k,f^k}$ is defined as $\Psi_{B,f}$, with $L_B$ replaced by
$L_{B_k}$ and $f_r$ replaced by
$f^k_r$. Thus we just have to prove that $\lim_k \E[\Psi
_{B_k,f^k}(V^k)]=\E[ \Psi_{B,f}(V)]$.
First, we know from Lemma~\ref{trws} that $L_{B}\phi$ is continuous on
$\rr^3\times\rr^3$. We deduce that $\Psi_{B,f}$ is continuous
at each $x\in\dd([0,\infty),\rr^3)$ such that $x$ has no jump at
$t_1,\dots,t_n,s,t$. But
$V$ has a.s. no jump at fixed points by (ii). Since $V^k$ goes in law
to $V$ and since
$f^k_r$ tends weakly to $f_r$ for each $r$ (because $V^k$ goes in law
to $V$ and since
$V$ has no fixed discontinuity),
we deduce that
$\Psi_{B,f^k}(V^k)$ goes in law to $\Psi_{B,f}(V)$.
Using that the family $(\Psi_{B,f^k}(V^k))_{k\geq1}$ is uniformly integrable
[because $\llvert \Psi_{B,f^k}(V^k)\rrvert  \leq C_\Psi(1+ \int_s^t \int_{\rr^3}
\llvert V^k_r-v\rrvert ^{\gamma+1}
f_r^k(dv)\,dr)\leq C_{t,\Psi} (1+\sup_{[0,t]} \llvert V^k_r\rrvert ^{\gamma+1})$ and
due to (i)],
we conclude that $\lim_k \E[\Psi_{B,f^k}(V^k)]=\E[ \Psi_{B,f}(V)]$.
Hence it only remains to check that $\lim_k \E[\llvert \Psi
_{B_k,f^k}(V^k)-\Psi
_{B,f^k}(V^k)\rrvert ]=0$.
Using point (i) and that $\llvert (L_B-L_{B_k})\phi(v,v_*)\rrvert \leq C_\phi
k^{-\kappa}(1+\llvert v\rrvert ^2+\llvert v_*\rrvert ^2)$
for some $\kappa>0$
(see the proof of Lemma~\ref{trws}), one easily concludes.
\end{pf}

We finally may give the following:

\begin{pf*}{Proof of Proposition~\ref{exifc}(\normalfont{ii})}
We thus assume
(\ref{aaa})
for some $\gamma\in(-1,0]$ and some $\nu\in(0,1)$ and consider
$f_0\in\cP_p(\rr^3)$ for some $p>2$. We know from Lemma~\ref{tight}
that there exists a solution $(V_t)_{t\geq0}$ to $\MP(f_0,B)$ and that
$\E[\sup_{[0,T]} \llvert V_t\rrvert ^p]\leq C_{T,p}$ for all $T>0$. For $t\geq0$, set
$f_t=\cL(V_t)$. Then (\ref{mom2}) obviously holds, since $m_p(f_t)=\E
[\llvert V_t\rrvert ^p]$.
Finally, Remark~\ref{eq} ensures us that $(V_t)_{t\geq0}$ solves
(\ref{bolsde})
and that $(f_t)_{t\geq0}$ is a weak solution to (\ref{be}) starting
from~$f_0$.
\end{pf*}

\subsection{Hard potentials}
We still have to prove Proposition~\ref{exifc}(i).
We use very similar arguments as in \cite{f:u}, Proof of Proposition~3.4,
concerning the $3D$ Boltzmann equation without cutoff
with velocity cross section $\min(\llvert v-v_*\rrvert ^\gamma, k)$.

In the whole subsection, we assume (\ref{aaa})
for some $\gamma\in(0,1)$, $\nu\in(0,1)$.
A~weak solution $(f_t)_{t\geq0}$ to (\ref{be}) starting from $f_0\in
\cP
_2(\rr^3)$
satisfying (\ref{mom}) is fixed.

For $t\geq0$, we introduce $A_t$ defined, for
$\phi\in\lip_b(\rr^3)$ and $v \in\rr^3$, by [recall (\ref{dfL}) and
(\ref{newL})]
%
\begin{eqnarray}
\label{at} A_t \phi(v) &=& \int_{\rr^3}L_B
\phi(v,v_*)f_t(dv_*)\nonumber
\\
&=&\int_{\rr^3}\int_0^{\pi/2}
\int_0^{2\pi} \llvert v-v_*\rrvert
^\gamma \\
&&\hspace*{61pt}{}\times\bigl[\phi\bigl(v+a(v,v_*,\theta,\varphi)\bigr)-\phi(v) \bigr]b(
\theta)\,d\varphi\, d\theta f_t(dv_*),
\nonumber
\end{eqnarray}
where $a$ was defined in (\ref{dfvprime}).
We define similarly, for $k\geq1$, setting $H_k(v)=\frac{\llvert v\rrvert \land
k}{\llvert v\rrvert } v$,
\begin{eqnarray*}
A_t^k \phi(v) &=& \int_{\rr^3}\int
_0^{\pi/2} \int_0^{2\pi}
\bigl\llvert H_k(v)-v_*\bigr\rrvert ^\gamma \\
&&\hspace*{59pt}{}\times\bigl[\phi
\bigl(v+a\bigl(H_k(v),v_*,\theta,\varphi\bigr)\bigr)-\phi(v) \bigr]
b(\theta )\,d\varphi\, d\theta f_t(dv_*).
\end{eqnarray*}

\begin{defi}(i) Let $t_0\geq0$ and $\mu\in\cP(\rr^3)$ be fixed.
A c\`adl\`ag
adapted process $(V_t)_{t\geq t_0}$ on some probability space $(\Omega
,\cF,(\cF_t)_{t\geq0},\Pr)$
solves the martingale problem $\MP(\mu,t_0,(A_t)_{t\geq t_0},C^1_c(\rr
^3))$ if
$\cL(V_{t_0})=\mu$ and if for all $\phi\in C^1_c(\rr^3)$,
$(M_t^\phi
)_{t\geq t_0}$
is a $(\Omega,\cF,(\cF_t)_{t\geq t_0},\Pr)$-martingale,\vspace*{2pt} where
$M^\phi_t:=\phi(V_t)-\int_{t_0}^t A_s \phi(V_s)\,ds$.

(ii) For $t_0\geq0$, $\mu\in\cP(\rr^3)$ and $k\geq1$, the
martingale problem
$\MP(\mu,t_0,\allowbreak(A_t^k)_{t\geq t_0},C^1_c(\rr^3))$ is defined similarly.
\end{defi}

The following remark is classical; see, for example, Tanaka \cite{t}, Section~4.

\begin{rk}\label{upp}
(i) A process $(V_t)_{t\geq t_0}$ on some probability space $(\Omega
,\cF
,\allowbreak(\cF_t)_{t\geq0},\Pr)$
is solution to $\MP(\mu,t_0,(A_t)_{t\geq t_0},C^1_c(\rr^3))$ if and only
if $\cL(V_{t_0})=\mu$ and if
there exists,
on a possibly enlarged probability space, a $(\cF_t)_{t\geq
0}$-Poisson measure
$N(ds,dv,d\theta,d\varphi,du)$ on
$[0,\infty)\times\rr^3 \times(0,\pi/2]\times[0,2\pi)\times
[0,\infty)$
with intensity $dsf_s(dv) b(\theta)\,d\theta\, d\varphi\, du$
such that for all $t\geq t_0$,
%
\begin{eqnarray}
\label{bolsde1} \hspace*{15pt}V_t&=&V_{t_0} + \int_{t_0}^t
\int_{\rr^3}\int_0^{\pi/2} \int
_0^{2\pi} \int_0^\infty
a(V_\sm,v,\theta,\varphi)\nonumber\\[-8pt]\\[-8pt]
\hspace*{15pt}&&\hspace*{122pt}{}\times\indiq_{\{u\leq\llvert V_\sm-v\rrvert ^\gamma\}
}N(ds,dv,d\theta,d
\varphi,du).\nonumber
\end{eqnarray}

(ii) Similarly, a process $(V_t^k)_{t\geq t_0}$ solves
$\MP(\mu,t_0,(A_t^k)_{t\geq t_0},C^1_c(\rr^3))$ if and only if $\cL
(V_{t_0})=\mu$ and if
it solves
%
\begin{eqnarray}
\label{bolsde2} \hspace*{15pt}V_t^k&=& V_{t_0} + \int
_{t_0}^t \int_{\rr^3}\int
_0^{\pi/2} \int_0^{2\pi}
\int_0^\infty a\bigl(H_k
\bigl(V_\sm^k\bigr),v,\theta,\varphi\bigr)\nonumber\\[-8pt]\\[-8pt]
\hspace*{15pt}&&\hspace*{122pt}{}\times
\indiq_{\{u\leq\llvert H_k(V_\sm
^k)-v\rrvert ^\gamma
\}}N(ds,dv,d\theta,d\varphi,du).\nonumber
\end{eqnarray}
\end{rk}

We start with the following statement.

\begin{rk}\label{euar}
For any $t_0\geq0$, any $\mu\in\cP_2(\rr^3)$ and any $k\geq1$,
there exists a unique (in law) solution $(V^k_t)_{t\geq t_0}$ to
$\MP(\mu,t_0,(A_t^k)_{t\geq t_0},C^1_c(\rr^3))$.
\end{rk}

This can be proved exactly as in \cite{f:u}, Proof of Proposition~3.4, Steps 1 to 7. We have checked
all the details and omit the proof. Let us only mention that we have to
use the following estimates:
(i) $\int_{\rr^3}f_s(dv_*) (\llvert H_k(v)-v_*\rrvert ^\gamma+\llvert H_k(v)-v_*\rrvert ^{\gamma+1})
\leq C_k$,
(ii) $\int_{\rr^3}f_s(dv_*) \llvert H_k(v)-v_*\rrvert ^\gamma|H_k(v)-H_k(\tilde
v)|\leq
C_k |v- \tilde v|$,
(iii)~$\int_{\rr^3}f_s(dv_*) |H_k(v)-v_*| ||H_k(v)-v_*|^{\gamma} -
|H_k(\tilde v)-v_*|^{\gamma}|
\leq C_k |v- \tilde v|$. Points (i) and (ii) are easily checked and use
only that $H_k
\in\lip_b(\rr^3)$ and that $\int_{\rr^3}f_s(dv_*) (1+|v_*|^\gamma
+|v_*|^{\gamma+1}) \leq
\int_{\rr^3}f_s(dv_*) (3+|v_*|^2)\leq C$ by (\ref{energy}). Point
(iii) uses
additionally (\ref{thf}).

To make tend $k$ to infinity, we will need the following uniform (in
$k$) moment estimates.

\begin{lem}\label{momk}
Consider the
solution $(V^k_t)_{t\geq t_0}$ to $\MP(\mu,t_0,(A_t^k)_{t\geq
t_0},C^1_c(\rr^3))$,
for some $t_0 >0$ and some $\mu\in\cP_2(\rr^3)$. For any $T>t_0$,
we have
\begin{longlist}[(ii)]
\item[(i)] $\sup_{[t_0,T]} \E[\llvert V_t^k\rrvert ^2]\leq C_{t_0,T,\mu}$,

\item[(ii)] $\E[\sup_{[t_0,T]} \llvert V_t^k\rrvert ]\leq C_{t_0,T,\mu}$.
\end{longlist}
\end{lem}

\begin{pf}
We start with (i). Using (\ref{bolsde2}), the It\^o formula for jump processes
(see, e.g., Jacod and Shiryaev \cite{js}, Theorem~4.57, page 56), taking
expectations and integrating in
$u$,
we get, for $t\geq t_0$,
\begin{eqnarray*}
\E\bigl[\bigl\llvert V_t^k\bigr\rrvert ^2
\bigr]&=&\E\bigl[\bigl\llvert V_{t_0}^k\bigr\rrvert
^2\bigr]\\
&&{}+\E \biggl[\int_{t_0}^t \int
_{\rr
^3}\int_0^{\pi
/2}\int
_0^{2\pi} \bigl(\bigl\llvert a
\bigl(H_k\bigl(V_s^k\bigr),v,\theta,\varphi
\bigr)\bigr\rrvert ^2 \\
&&\hspace*{106pt}{}+2 \bigl\langle V^k_s,
a\bigl(H_k\bigl(V_s^k\bigr),v,\theta,
\varphi\bigr) \bigr\rangle \bigr)
\\
&&\hspace*{101pt}{}\times \bigl\llvert H_k\bigl(V^k_s\bigr)-v\bigr
\rrvert ^\gamma b(\theta)\,d\varphi\, d\theta f_s(dv)\,ds
\biggr].
\end{eqnarray*}
After some explicit computation using (\ref{dfvprime}) and (\ref{lit}),
this yields
\begin{eqnarray*}
\E\bigl[\bigl\llvert V_t^k\bigr\rrvert ^2
\bigr]&=&\int_{\rr^3}\llvert v\rrvert ^2\mu(dx) \\
&&{}+ \E
\biggl[\int_{t_0}^t \int_{\rr^3}
\int_0^{\pi/2} \bigl(\bigl\llvert H_k
\bigl(V_s^k\bigr)-v\bigr\rrvert ^2- 2 \bigl
\langle V^k_s, H_k\bigl(V_s^k
\bigr)-v \bigr\rangle \bigr)
\\
&&\hspace*{80pt}{}\times \pi\bigl\llvert H_k\bigl(V^k_s\bigr)-v
\bigr\rrvert ^\gamma(1-\cos\theta)b(\theta)\,d\theta f_s(dv)
\,ds \biggr].
\end{eqnarray*}
Observe that $(1-\cos\theta)b(\theta)$ is integrable due to
(\ref{aaa}).
Next, we have $ \langle V^k_s,\allowbreak H_k(V_s^k) \rangle\geq
\llvert H_k(V_s^k)\rrvert ^2$ and
$\llvert H_k(V^k_s)\rrvert \leq\llvert V^k_s\rrvert $,
from which we deduce\vspace*{2pt}
$|H_k(V_s^k)-v|^2- 2 \langle V^k_s, H_k(V_s^k)-v \rangle
\leq|v|^2 + 2  \langle
V^k_s - H_k(V_s^k),v  \rangle
\leq|v|^2 + 2 |V^k_s| |v|$. We also have $|H_k(V^k_s)-v|^\gamma
\leq C(1+|H_k(V^k_s)|+|v|)\leq C (1+|V^k_s|+|v|)$. We finally find
that\vspace*{1pt}
$(|H_k(V_s^k)-v|^2- 2 \langle V^k_s, H_k(V_s^k)-v \rangle
)|H_k(V^k_s)-v|^\gamma
\leq
C(|v|^2 + |V^k_s| |v|)(1+|V^k_s|+|v|)\leq C (1+|v|^3)(1+|V^k_s|^2)$. Thus
\begin{eqnarray*}
\E\bigl[\bigl\llvert V_t^k\bigr\rrvert ^2
\bigr]&\leq& C_\mu+ C \E \biggl[\int_{t_0}^t
\int_{\rr^3}\bigl(1+\llvert v\rrvert ^3\bigr)
\bigl(1+\bigl\llvert V^k_s\bigr\rrvert ^2
\bigr) f_s(dv)\,ds \biggr] \\
&\leq& C_\mu+ C_{t_0}
\int_{t_0}^t \E\bigl[1+\bigl\llvert
V^k_s\bigr\rrvert ^2\bigr]\,ds.
\end{eqnarray*}
We used that, since $t_0>0$, $\sup_{t\geq t_0} m_3(f_s)<\infty$ by
(\ref{mom}).
The Gr\"onwall lemma thus implies $\sup_{[t_0,T]}\E[\llvert V_t^k\rrvert ^2]\leq
C_{t_0,T,\mu}$ as desired.

Point (ii) easily follows, since
\begin{eqnarray*}
\E \Bigl[\sup_{[t_0,T]}\bigl\llvert V_s^k
\bigr\rrvert \Bigr]&\leq&\E\bigl[\bigl\llvert V_{t_0}^k\bigr
\rrvert \bigr]\\
&&{}+\E \biggl[\int_{t_0}^T \int
_{\rr^3}\int_0^{\pi/2}\int
_0^{2\pi} \bigl\llvert a\bigl(H_k
\bigl(V_s^k\bigr),v,\theta,\varphi\bigr)\bigr\rrvert
\bigl\llvert H_k\bigl(V^k_s\bigr)-v\bigr
\rrvert ^\gamma
\\
&&\hspace*{158pt}{}\times b(\theta)\,d\varphi \,d\theta f_s(dv)\,ds \biggr],
\end{eqnarray*}
so that using (\ref{lit}) and that $\theta b(\theta)$ is integrable by
(\ref{aaa}),
\begin{eqnarray*}
\E \Bigl[\sup_{[t_0,T]}\bigl\llvert V_s^k
\bigr\rrvert \Bigr]&\leq&\int_{\rr^3}\llvert v\rrvert \mu(dv) + C
\E \biggl[\int_{t_0}^T \int_{\rr^3}
\bigl\llvert H_k\bigl(V^k_s\bigr)-v\bigr
\rrvert ^{\gamma+1} f_s(dv)\,ds \biggr]
\\
&\leq& C_\mu+ C \int_{t_0}^T \int
_{\rr^3}\bigl(1+ \E\bigl[\bigl\llvert V^k_s
\bigr\rrvert ^2\bigr] + \llvert v\rrvert ^2\bigr)
f_s(dv)\,ds \leq C_{t_0,T,\mu}
\end{eqnarray*}
by (i) and (\ref{energy}).
\end{pf}

We deduce the well-posedness of $\MP(\mu,t_0,(A_t)_{t\geq
t_0},C^1_c(\rr
^3))$ when $t_0>0$.

\begin{lem}\label{unitmp}
Let $t_0 >0$ and $\mu\in\cP_2(\rr^3)$ be fixed.
There exists a unique (in law) solution $(V_t)_{t\geq t_0}$ to
$\MP(\mu,t_0,(A_t)_{t\geq t_0},C^1_c(\rr^3))$.
\end{lem}

\begin{pf} We only sketch the proof, since it is tedious but rather standard.

\textit{Uniqueness.} Consider $(V_t)_{t\geq t_0}$ solving $\MP(\mu
,t_0,(A_t)_{t\geq t_0},C^1_c(\rr^3))$. Introduce,
for $k\geq1$, $\tau_k=\inf\{t\geq t_0 \dvtx  \llvert V_t\rrvert \geq k\}$ (with the convention
that $\tau_k=t_0$ if this set is empty). Since $(V_t)_{t\geq t_0}$ is
c\`adl\`ag by assumption,
it is locally bounded, whence $\tau_k \to\infty$ a.s. as $k \to
\infty$.
For $k\geq1$, observe that $V$ solves $\MP(\mu,t_0,(A_t^k)_{t\geq
t_0},C^1_c(\rr^3))$ until
$\tau_k$ (because $v=H_k(v)$ if $\llvert v\rrvert \leq k$ and because $\llvert V_t\rrvert  <k$ for
all $t\in[t_0,\tau_k)$).
By uniqueness for $\MP(\mu,t_0,(A_t^k)_{t\geq t_0},\break C^1_c(\rr^3))$, we
deduce that for any $T>0$, any $k\geq1$,
the law of $(V_t)_{t\in[t_0,T]}$ knowing $\tau_k>T$ is entirely determined.
Using that $\tau_k\to\infty$ a.s. as $k\to\infty$, we easily conclude.

\textit{Existence.} One way to prove such an existence result is to use a
tightness argument
as in Lemma~\ref{tight} above. Another way is the following. Consider
$T>t_0$ arbitrarily large.
Roughly, if $k$ is very large, then
a solution $(V^k_t)_{t\geq t_0}$ to $\MP(\mu,t_0,(A_t^k)_{t\geq
t_0},C^1_c(\rr^3))$
will not reach $k$ before $T$ with a high probability [due to Lemma~\ref
{momk}(ii)],
so that it actually also solves $\MP(\mu,t_0,(A_t)_{t\geq
t_0},\break C^1_c(\rr
^3))$ during $[t_0,T]$
[because as previously, $v=H_k(v)$ for $|v|\leq k$].
\end{pf}

The last preliminary will be useful to show that the law of $V_t$ is
indeed~$f_t$.

\begin{lem}\label{unilbe}
Let $t_0 >0$ and $\mu\in\cP(\rr^3)$ be fixed.
There exists at most one family $(\mu_t)_{t\geq0}\subset\cP(\rr^3)$
such that for all
$\phi\in C^1_c(\rr^3)$, all $t\geq t_0$,
\[
\int_{\rr^3}\phi(v) \mu_t (dv)= \int
_{\rr^3}\phi(v) \mu(dv) + \int_{t_0}^t
\int_{\rr^3}A_s \phi(v)\mu_s(dv)\,ds.
\]
\end{lem}

\begin{pf} This will follow from Horowitz and Karandikar \cite{hk}, Theorem B1, if we check the
following points:
\begin{longlist}[(b)]
\item[(a)] $C^1_c(\rr^3)$ is dense in $C_0(\rr^3)$ for the uniform convergence
topology;

\item[(b)] $(t,v)\mapsto A_t\phi(v)$ is measurable for all $\phi\in
C^1_c(\rr^3)$;

\item[(c)] for each $t\geq0$, $A_t$ satisfies the maximum principle;

\item[(d)] there exists a countable subset $\{\phi_k\}\subset C^1_c(\rr^3)$
such that for all $t\geq t_0$,
the closure of $\{(\phi_k,A_t\phi_k) \dvtx  k\geq1\}\subset C^1_c(\rr
^3)$ for
the bounded-pointwise convergence
is $\{(\phi,A_t\phi) \dvtx  \phi\in C^1_c(\rr^3)\}$;

\item[(e)] for all $v_0\in\rr^3$, $\MP(\delta_{v_0},t_0,(A_t)_{t\geq
t_0},C^1_c(\rr^3))$ is well posed.
\end{longlist}

First, (a) and (b) are clear, and (e) follows from Lemma~\ref{unitmp}.
Next, (c) is obvious
from (\ref{at}): if $\phi$ attains its maximum at some
$v_0 \in\rr^3$, $A_t\phi(v_0)\leq0$. The only delicate point is (d).
Consider a countable family $\{\phi_k\}_{k\geq1}\subset
C^1_c(\rr^3)$ dense in $C^1_c(\rr^3)$ in the following sense: for all
$\phi\in C^1_c(\rr^3)$ such that
$\supp\phi\subset\cB(0,R)$, there is a subsequence $\phi_{k_n}$ such
that $\supp\phi_{k_n}
\subset\cB(0,R+1)$
and $\llVert  \phi- \phi_{k_n}\rrVert _{L^\infty(\rr^3)} +\llVert  \nabla(\phi-
\phi
_{k_n})\rrVert _{L^\infty(\rr^3)} \to0$.
We have to prove that $(\phi_{k_n},A_t\phi_{k_n})$ goes to $(\phi
,A_t\phi)$ bounded-pointwise.
We obviously have that $\phi_{k_n} \to\phi$ bounded-pointwise. An
immediate computation using
(\ref{lit}), (\ref{aaa})
and (\ref{energy})
shows that for all $v\in\rr^3$, $\llvert A_t\phi_{k_n}(v)-A_t\phi(v)\rrvert  \leq C
\llVert  \nabla(\phi- \phi_{k_n})\rrVert _{L^\infty(\rr^3)} \int_{\rr
^3}\theta
\llvert v-v_*\rrvert ^{\gamma+1} b(\theta)\,d\theta f_t(dv_*)
\leq C\llVert  \nabla(\phi- \phi_{k_n})\rrVert _{L^\infty(\rr^3)} (1+\llvert v\rrvert ^2)
\to
0$. It only remains to prove that
$\sup_{v\in\rr^3} \sup_{n \geq1} \llvert A_t\phi_{k_n}(v)\rrvert <\infty$.

To this end, it suffices to check that
for $\phi\in C^1_c(\rr^3)$ with $\llVert \phi\rrVert _{L^\infty(\rr^3)}+
\llVert \nabla
\phi\rrVert _{L^\infty(\rr^3)} \leq K$
and $\supp\phi\subset\cB(0,R)$, we have $\llVert A_t\phi\rrVert _{L^\infty
(\rr
^3)} \leq C_{K,R}$.

First consider $v \in\rr^3$ such that $\llvert v\rrvert  \leq5R$. Then using (\ref
{lit}), (\ref{aaa})
and (\ref{energy}), we obtain
$\llvert A_t\phi(v)\rrvert \leq K \int_{\rr^3}\theta\llvert v-v_*\rrvert ^{\gamma+1} b(\theta)\,d\theta
f_t(dv_*) \leq
C K (1+\break \llvert v\rrvert ^{\gamma+1}) \leq C K ( 1+R^{\gamma+1})$.

Next, consider $v\in\rr^3$ such that $\llvert v\rrvert  \geq5R$. Then we have
$\phi
(v)=0$, so that
$\llvert \phi(v+a(v,v_*,\theta,\varphi))-\phi(v)\rrvert  \leq K \llvert a(v,v_*,\theta
,\varphi)\rrvert
\indiq_{\{\llvert v+a(v,v_*,\theta,\varphi)\rrvert  < R\}}$. But
$\llvert v+a(v,v_*,\theta,\varphi)\rrvert  < R$ implies $\llvert a(v,v_*,\theta,\varphi
)\rrvert  >
|v|-R\geq4|v|/5$,
whence [recall (\ref{lit})]
$\sqrt{1-\cos\theta} |v-v_*|>4\sqrt2 |v|/5$, from which (recall that
$\theta\in(0,\pi/2]$)
$|v|+|v_*| > 4\sqrt2|v|/5$ and finally $|v_*| > (4\sqrt2 / 5 - 1)|v|
> |v|/10$.
We thus get $|\phi(v+a(v,v_*,\theta,\varphi))-\phi(v)| \leq
K|a(v,v_*,\theta,\varphi)| \indiq_{\{ |v_*|>|v|/10\}}
\leq K \theta|v-v_*|\* \indiq_{\{ |v_*|>|v|/10\}}$ by (\ref{lit}), whence
\begin{eqnarray*}
\bigl\llvert A_t\phi(v)\bigr\rrvert \leq K \int
_{\rr^3}\int_0^{\pi/2}\int
_0^{2\pi} \theta \llvert v-v_*\rrvert
^{1+\gamma}\indiq_{\{ \llvert v_*\rrvert >\llvert v\rrvert /10\}} b(\theta)\,d\theta\, d\varphi
f_t(dv_*).
\end{eqnarray*}
Using (\ref{aaa}) and then (\ref{energy}), we deduce that
\begin{eqnarray*}
\bigl\llvert A_t\phi(v)\bigr\rrvert &\leq& K \int_{\rr^3}
\llvert v-v_*\rrvert ^{1+\gamma}\indiq_{\{
\llvert v_*\rrvert >\llvert v\rrvert /10\}
}f_t(dv_*)\\
&\leq& K \int_{\rr^3}\bigl(11\llvert v_*\rrvert
\bigr)^{\gamma+1} f_t(dv_*) \leq CK. %
\end{eqnarray*}
We finally have checked that for any $v\in\rr^3$, $\llvert A_t\phi(v)\rrvert \leq C
K(1+R^{\gamma+1})$.
\end{pf}

We finally may give the

\begin{pf*}{Proof of Proposition~\ref{exifc}(\normalfont{i})} We
divide the
proof into two steps.

\textit{Step 1.} For $t_0>0$, let $(V_t)_{t\geq t_0}$ be the unique (in
law) solution to
$\MP(f_{t_0},t_0,\allowbreak(A_t)_{t\geq t_0},C^1_c(\rr^3))$. The aim of this step
is to prove that $\cL(V_t)=f_t$ for
all $t\geq t_0$. To this end, put $\mu_t=\cL(V_t)$.
For any $\phi\in C^1_c(\rr^3)$ and any $t\geq t_0$, we know that
$\phi
(V_t)-\int_{t_0}^t A_s\phi(V_s)\,ds$
is a martingale, whence $\E[\phi(V_t)-\int_{t_0}^t A_s\phi
(V_s)\,ds]=\E
[\phi(V_{t_0})]$, which yields
\[
\int_{\rr^3}\phi(v) \mu_t(dv) = \int
_{\rr^3}\phi(v) f_{t_0}(dv) + \int
_{t_0}^t \int_{\rr^3}A_s
\phi(v)\mu_s(dv)\,ds. %
\]
But $(f_t)_{t\geq0}$ is a weak solution to (\ref{be}), whence,
for $\phi\in C^1_c(\rr^3)\subset\lip_b(\rr^3)$ and $t\geq t_0$,
\begin{eqnarray*}
\int_{\rr^3}\phi(v) f_t(dv) &=& \int
_{\rr^3}\phi(v) f_{t_0}(dv) + \int
_{t_0}^t \int_{\rr^3}\int
_{\rr^3}L_B \phi(v,v_*)f_s(dv_*)
f_s(dv)\,ds
\\
&=& \int_{\rr^3}\phi(v) f_{t_0}(dv) + \int
_{t_0}^t \int_{\rr
^3}A_s
\phi (v)f_s(dv)\,ds.
\end{eqnarray*}
Lemma~\ref{unilbe} implies that $\mu_t=f_t$ for all $t\geq t_0$.

\textit{Step 2.} We deduce from Step 1 that if $(V_t^{t_0})_{t\geq t_0}$ solves
$\MP(f_{t_0},t_0,(A_t)_{t\geq t_0},\allowbreak C^1_c(\rr^3))$, then for any
$t_1>t_0$, $(V_t^{t_0})_{t\geq t_1}$ solves
$\MP(f_{t_1},t_1,(A_t)_{t\geq t_1},C^1_c(\rr^3))$.
This compatibility property [recall that uniqueness holds for
$\MP(f_{t_0},t_0,\break (A_t)_{t\geq t_0},\allowbreak C^1_c(\rr^3))$ for any $t_0>0$ by
Lemma~\ref{unitmp}] implies,
by the Kolmogorov theorem, that there exists a process $(V_t)_{t\geq
0}$ such that for all
$t_0>0$, $(V_t)_{t\geq t_0}$ solves
$\MP(f_{t_0},t_0,(A_t)_{t\geq t_0},C^1_c(\rr^3))$. In particular, we
have $\cL(V_t)=f_t$ for all $t>0$ by Step 1.
Since now $f_{t_0}$ tends weakly to $f_0$ as $t_0\to0$ [use, e.g.,
Lemma \ref{trws}],
we easily deduce that $(V_t)_{t\geq0}$ solves $\MP(f_{0},0,(A_t)_{t\geq
0},C^1_c(\rr^3))$. Due to Remark~\ref{upp}(i), this ends the proof.
\end{pf*}




\printaddresses

\end{document}